\documentclass[aps,amsmath,amssymb,pra,longbibliography,showpacs,showkeys,twocolumn,floatfix]{revtex4-2}
\usepackage{graphicx}
\usepackage{dcolumn}
\usepackage{bm}
\usepackage[dvipsnames,svgnames,x11names]{xcolor}
\usepackage{natbib}
\usepackage{hyperref}
\usepackage[mathlines]{lineno}
\usepackage{listings}
\usepackage{subcaption}
\usepackage{xcolor}
\usepackage[collision]{chemsym}

\begin{document}


\title{L-shell Photoionisation Cross Sections in the \S$^+$, \S$^{2+}$, \S$^{3+}$ Isonuclear Sequence}



\author{J.-P. Mosnier}
\email[corresponding author ]{jean-paul.mosnier@dcu.ie}
\affiliation{School of Physical Sciences, National Centre for Plasma Science and Technology (NCPST), Dublin City University, Glasnevin, Dublin 9, Ireland}

\author{E.T. Kennedy}
\email{eugene.kennedy@dcu.ie}
\affiliation{School of Physical Sciences, National Centre for Plasma Science and Technology (NCPST), Dublin City University, Glasnevin, Dublin 9, Ireland}

\author{D. Cubaynes}
\email{denis.cubaynes@universite-paris-saclay.fr}
\affiliation{Institut des Sciences Mol\'{e}culaires d’Orsay, UMR 8214, Rue Andr\'{e} Rivière, Bâtiment 520, Universit\'{e} Paris-Saclay, 91405 Orsay, France \\ and \\ Synchrotron SOLEIL, L’Orme des Merisiers, Saint-Aubin, BP 48, CEDEX, F-91192 Gif-sur-Yvette, France}

\author{J.-M. Bizau}
\email{bizau.jean-marc@orange.fr }
\affiliation{Institut des Sciences Mol\'{e}culaires d’Orsay, UMR 8214, Rue Andr\'{e} Rivière, Bâtiment 520, Universit\'{e} Paris-Saclay, 91405 Orsay, France \\ and \\ Synchrotron SOLEIL, L’Orme des Merisiers, Saint-Aubin, BP 48, CEDEX, F-91192 Gif-sur-Yvette, France}

\author{S. Guilbaud}
\email{segolene.guilbaud@universite-paris-saclay.fr}
\affiliation{Institut des Sciences Mol\'{e}culaires d’Orsay, UMR 8214, Rue Andr\'{e} Rivi\`{e}re, Bâtiment 520, Universit\'{e} Paris-Saclay, 91405 Orsay, France}

\author{C. Blancard}
\email{christophe.blancard@cea.fr}
\affiliation{CEA, DAM, DIF, F-91297 Arpajon, France \\ and \\
Université Paris-Saclay, CEA, LMCE, F-91680 Bruyères-le-Châtel, France}

\author{B.M. McLaughlin}
\affiliation{Centre for Theoretical Atomic, Molecular and Optical Physics, School of Mathematics and Physics, Queen's University Belfast, BT7 1NN, UK}

\author{M.F. Haso\u{g}lu}
\email{mfatih.hasoglu@hku.edu.tr }
\affiliation{Aeronautics and Aerospace Engineering, Hasan Kalyoncu University, Gaziantep, Turkey}

\author{T.W. Gorczyca}
\email{thomas.gorczyca@wmich.edu }
\affiliation{Department of Physics, Western Michigan University, Kalamazoo, MI, 49008-5200 USA}


\date{\today}

\begin{abstract}
We present absolute L-shell photoionisation cross sections for the \S$^{+}$, \S$^{2+}$, \S$^{3+}$ ions. The cross sections were obtained using the monochromatised photon beam delivered by the SOLEIL synchrotron source coupled with an ion beam extracted from an electron cyclotron resonance source (ECRIS) in the merged dual-beam configuration. The cross sections for single, double and triple ionisation were measured and combined to generate total photoionisation cross sections. For each of the \S$^{+}$, \S$^{2+}$ and \S$^{3+}$ ions, the photon energy regions corresponding to the excitation and ionisation of a 2p or a 2s electron ($\sim$ 175-230 eV) were investigated.
The experimental results are interpreted with the help of multiconfigurational Dirac-Fock (MCDF) and Breit-Pauli R-Matrix (BPRM) or Dirac R-Matrix (DARC) theoretical calculations. The former generates photoabsorption cross sections from eigenenergies and eigenfunctions obtained by solving variationally the multiconfiguration Dirac Hamiltonian while the latter calculate cross sections for photon scattering by atoms. 
The cross sectional spectra feature rich resonance structures with narrow natural widths (typically $\le$ 100 meV) due to ${\textnormal{2p}}\rightarrow n{\textnormal{d}}$ excitations below and up to the  ${\textnormal{2p}}$ thresholds. This behaviour is consistent with the large number of inner-shell states based on correlation and spin-orbit mixed configurations having three open subshells. Strong and wide (typically $\sim$ 1 eV) Rydberg series of resonances due to ${\textnormal{2s}}\rightarrow n{\textnormal{p}}$ excitations dominate above the  ${\textnormal{2p}}$ threshold.
\end{abstract}


\maketitle





\section{\label{Intro}Introduction}
Sulphur $({\textnormal{Z}}=16)$ is ubiquitous in the universe/cosmos (tenth most abundant chemical element by mass, \S/\H  abundance of $1.4\times10^{-5}$, fifth most abundant on Earth)  where it is found in many gaseous (free) or condensed atomic and molecular states. Importantly, sulfur enters the chemical composition of many of the essential proteins found in living organisms due to its presence in the methionine and cysteine amino-acids building blocks. The understanding of the fundamental interactions occurring in short wavelength photon absorption/scattering by neutral or charged sulfur atomic and molecular species is key to (a) unravelling the structure and dynamics of complex sulfurated astrophysical environments, see e.g.,\citep{Adamkovics_2011,Frankel_2009,Fuente_2016,HilyBlant_2022,Mifsud_2021} and (b) interpreting XPS data to gain insight into the molecular shapes and valency of sulfur based biomolecules, see e.g.,\citep{Artemenko_2021,Marti_2004}  

Singly and multiply charged sulphur atomic ions have been widely observed in, e.g, planetary nebulae and near high-energy regions such as hot massive stars and supernova shock fronts. The Hubble telescope filters selected a transition associated with singly ionised sulphur S$^+$ ions to image such regions. Charge exchange collisions between fully stripped sulphur ions (S$^{16+}$) explained an astrophysical 3.5 keV X-ray observation, which was initially considered a possible dark matter signal \citep{Shah_2016}. The Chandra \citep{Chandra} and XMM-Newton \citep{XMM-Newton} short wavelength observatories, in particular, have provided spectrally resolved data on a wide range of astrophysical environments where ionised species, including sulphur ions, play important roles. The recently launched XRISM \citep{xrism} and the future Athena missions \citep{Athena}will provide even greater sensitivity and higher spectral resolution data, requiring concomitant advances in the experimental measurement and theoretical calculations of key atomic data. Such laboratory astrophysics \citep{Kalman_2007,Savin_2012,Dubernet_2016,Nahar_2020,Muller_2017,Schippers_2020} investigations are essential to benchmark the different calculational approaches to providing the wide range of atomic data required for the interpretation of astrophysical phenomena. The interaction of ionising photons with free ions becomes particularly significant in the vicinity of strong sources of short wavelength radiation and modelling of such radiation dominated plasma regions requires detailed knowledge of the relevant photoionisation interactions, including data on transition energies and associated cross sections \citep{Foster_2010}. Photoionisation studies also provide insight into the inverse atomic process of dielectronic recombination, which can often play an important role in the equilibrium behaviour of astrophysical plasma environments \citep{Savin_2002}.

The interactions of ionising photons with several sulphur ion stages have been previously studied, particularly in the photon energy regions corresponding to valence shell ionisation. Singly charged \S$^{+}$ ions were investigated both theoretically \citep{Tayal_2006} and experimentally \citep{Kristensen_2002} and theoretical calculations were carried out for \S$^{2+}$ \citep{Nahar_1993}. The Breit-Pauli theoretical method was used to investigate triply ionised \S$^{3+}$ \citep{Stancalie_2018}. Theoretical calculation of oscillator strengths and photoionization cross sections were carried out along the magnesium sequence (including \S$^{4+}$) \citep{Butler_1984,Serrao_1995,Kim_2014} and a combined experimental and theoretical study \citep{Mosnier_2022}.

In this work, we measure the cross sections for the photo-excitation and photo-ionisation of an inner-shell 2p or a 2s electron (175-250 eV photon energy range) in the isonuclear series \S$^{+}$, \S$^{2+}$ and \S$^{3+}$. These atomic ions and their associated inner-shell photon interactions are of astrophysical interest \citep{Gatuzz_2023,Psaradaki_2024} and useful for the characterisation of the electronic structure of condensed sulfurated compounds in which the sulfur is in a low oxidation state. Photoionisation studies along isonuclear sequences, in which the same single electron  $nl \rightarrow m(l\pm1)$ dipole excitation takes place in an ion where one outer electron is systematically removed, have proven powerful to reveal new fundamental insights in atomic physics, see e.g., the original seminal work on \Ba, \Ba$^+$, \Ba$^{2+}$ \citep{Lucatorto_1981} and the body of work on the $4f$ collapse in xenon ions, e.g. \citep{Bizau_2006} and references therein. Due to significant astrophysical interest, photoionisation works along iron isonuclear sequences abound, e.g. \citep{ElHassan_2009,Gharaibeh_2011}. In laser-produced and other laboratory plasmas, a mixture of populations of positive ions with charge states forming short isonuclear (e.g. $z-1, z, z+1$) series typically determine the total plasma opacity. This will require the knowledge of the photoionisation cross sections of each of the $z-1, z, z+1$ individual ions \citep{Blancard_2017}. 

The \S$^{+}$, \S$^{2+}$ and \S$^{3+}$ ions belong respectively to the phosphorus-, silicon- and aluminium isoelectronic sequences. To our knowledge, the only L-shell photoionisation works carried out previously in any of the early members (2nd Row of Periodic Table) of these sequences are the laser-plasma based photoabsorption work of ~\citep{Costello_1992} and the photo-ion (absolute cross sections) work of \citep{Kennedy_2014}, both in \Si$^+$ (aluminium sequence). There is also the work of Kristensen {\it{et al.}}~\citep{Kristensen_2002} on \S$^{+}$ which we discuss in Section~\ref{P-like} below.

$2p$ and $2s$ electron excitations in the \S$^{+}$, \S$^{2+}$ and \S$^{3+}$ ions lead to several open subshell configurations producing multiple $LSJ$ states via angular momentum recoupling of the parents and grand-parents terms. The non-radiative decay of these states will be seen as a complex structure featuring multiple resonances in the photo-ion yield spectra. The results of advanced atomic structure/dynamics computer codes are needed to interpret the experimental data. Convergence between the measured and calculated data ultimately validates the theoretical framework used for the calculations. Here, the absolute data are compared with theoretical cross sections calculated within two different frameworks, namely the R-matrix and the multiconfiguration Dirac-Fock methods. This approach has proved successful for our recent work\citep{Mosnier_2022,Mosnier_2023} in the magnesium sequence, for example. 

\section{\label{expt}Experimental Details}

A detailed account of the MAIA (Multi-Analysis Ion Apparatus) apparatus and the general experimental procedures used in this merged-beam work will be found in ~\citep{Bizau_2016}. In a merged-beam experiment,  the photoionisation cross section value $\sigma(E)$ is obtained from the following equation:

\begin{equation} \label{xsection}
\sigma(E)=\frac{S(E)e^2\eta vq}{IJ\epsilon\int_0^L\frac{dz}{\Delta x \Delta yF_{xy}(z)}}
\end{equation}

\begin{table*}
\caption{\label{Expar}Sample values of main experimental parameters in the single ionisation and double ionisation channels of  \S$^{+}$, \S$^{2+}$ and \S$^{3+}$ at the $2p \rightarrow 3d$ resonance energies of 176.7, 181.4 and 186.3 eV, respectively.}
\begin{ruledtabular}
\begin{tabular}{lccc}
& \S$^{+}$  & \S$^{2+}$  & \S$^{3+}$  \\
\cline{2-4} 
Photoion count SI/DI (s$^{-1}$) & 65,400/10,500 & 10500/1820 & 5275/46 \\
Background count SI/DI (s$^{-1}$)& 12,500/9 & 9/130 & 310/12 \\
Photodiode current SI/DI $I$($\mu$A) & 151/147 & 149.5/152 & 141/141 \\
Ion current SI/DI $J$(nA) & 220/206 & 170/108 & 244/155 \\
Ion beam velocity $v$ (ms$^{-1})$ & $1.75\times10^{5}$ & $2.21\times10^{5}$ & $3.03\times10^{5}$ \\
Tag voltage (kV) &  -1.0   &   -1.0  &  -1.0    \\
ECRIS RF power (W) &  $10^{-4}$    &    10     &       6     \\
Channel plates efficiency $\epsilon$ & 0.56 & 0.56 & 0.56 \\
Form factor \footnote{This is the value of the integral in eqn~\ref{xsection}} (m$^{-1}$)   &   52,000    &    54,000 &  32,000
\end{tabular}
\end{ruledtabular}
 \end{table*}

 where $S(E)$ is the counting rate of the photo-ions produced by the synchrotron photons of energy $E$, $e$ is the charge of the electron, $\eta$ is the efficiency of the photodiode used to characterise the photon beam, $v$ is the magnitude  of the ions velocity in the interaction region, $q$ is the charge on the ion, $I$ is the current produced by the calibrated photodiode, $J$ is the ion current, $\epsilon$ is the efficiency of the channel plates used to measure the photo-ions. The integral takes account of the beam overlap geometry, with $z$ the ion beam propagation axis, $F_{xy}$ is the reduced form factor and $\Delta x\,\text{and}\, \Delta y$ are scan step parameters to obtain the ion beam profile along the $x$ and $y$ directions, respectively. The specific experimental details for the present measurements are given below or in Table~\ref{Expar} as follows. The \S$^{+}$, \S$^{2+}$ and \S$^{3+}$ sample atomic ions were produced from ${\H_2\S}$ gas, in a permanent magnet electron cyclotron resonance ion source (ECRIS) excited by a 12.36 GHz microwave power supply,  operated at the powers indicated in Table~\ref{Expar}. For each of the photoionisation experiments, the individual beam of \S$^{+}$, or \S$^{2+}$ or \S$^{3+}$ ions was extracted and accelerated through a potential difference of $-4$ kV to its respective terminal velocity $v_{\text{4kV}} \ \text{ms}^{-1}$ . A  magnetic filter and electrostatic deflector, tuned to $v_{\text{4kV}}\,  (\text{ms}^{-1)}$, then selected and guided the sample ion beam to a spatially well-defined interaction region (length $L=0.57$ m) where it merged with the counter-propagating synchrotron radiation (SR) beam. The left-handed circularly polarised synchrotron light from an undulator and a 600 l/mm grating was used for all the experiments. The background pressure in the interaction zone was $ \simeq 2\times10^{-9}$ mbar in all the experiments. Typical \S$^{+}$, \S$^{2+}$ and \S$^{3+}$ parent ion beam currents are indicated in Table~\ref{Expar}. After the interaction region, the remaining parent ions were collected in a Faraday cup while the photoion yields were counted with a microchannel plate. Photon energy scans of the photo-ions count rates were carried out to map out the energy dependence of the  photoionisation cross sections. By measuring the photon and ion beam parameters, their overlap volumes (see equation~\ref{xsection}) and using calibrated photon and ion detectors it was possible to obtain the measured cross sections on an absolute basis after noise subtraction \citep{Bizau_2016}. Suitable high-voltage biases (Tag voltages in Table~\ref{Expar}) were applied to the interaction region in order to distinguish the photoions produced within the interaction region by their value of $v \ \text{ms}^{-1}$ (eqn~\ref{xsection}) from those produced outside the region. 
 
 The photon energies were calibrated using a gas cell and known argon reference lines \citep{Ren_2011}. The resonance energies could be determined to within an accuracy of 40 meV. The relative uncertainty in the measured absolute cross sections is generally within 15\%. Reference \citep{Bizau_2016} provides further insight into the determination of the experimental uncertainties. Reliable experimental absolute cross sections are very important as uncertainties in atomic data can significantly affect, for example, the determination of chemical abundances \citep{Luridiana_2011}
or photoionisation modelling of astrophysical plasmas \citep{Ballhausen_2023}.  As \S$^{2+}$ and \O$^{+}$ have equal $q/m$ ratios, the natural contamination of the \S$^{2+}$ parent beam by \O$^{+}$ was determined from the measurement of the oxygen ions resonances in the 528-534 eV photon region (\citep{McLaughlin_2014,Bizau_2015}). Finally, we note that the same apparatus was used in quite similar experimental conditions to measure the photoionisation of \S$^{4+}$ \citep{Mosnier_2022}, however, the conditions proved unfavourably noisy to generate suitable beams of the sodium-like \S$^{5+}$ ion.


It is well known that populations of metastable states can be produced in the ECRIS source \citep{Kronholm_2018} in concentrations that depend on experimental conditions and are, therefore, variable. The measured cross sections will thus comprise weighted contributions from ions in the ground state (GS) and in the metastable state(s) (MS). Thus, accurate modelling of the experimental data from the results of theoretical cross section calculations must include both the GS and MS contributions in relative proportions. Relevant energy level data for the \S$^{+}$, \S$^{2+}$ and \S$^{3+}$ atomic ions is given in Table~\ref{Excitation Schemes}, and more complete information can be found in~\citep{Kramida_2023}.


\section{\label{theo}Theoretical Aspects}
	\subsection{\label{overview}Overview of electronic excitation and decay schemes}
	In the present photoion yield experiments, singly (SI), doubly (DI) and triply (TI) ionised species (the photoions) are collected following the absorption of a single photon ($h\nu$) by a singly (\S$^{+}$), doubly (\S$^{2+}$) or triply (\S$^{3+}$) charged sulphur ion. This can be symbolised by the following set of reactions, with $m=1,2 \ \text{or} \ 3$ and $\bar{e}$ a free electron, :
		\begin{align*}
		\S^{m+}+h\nu \longrightarrow \ &\S^{(m+1)}+ \ 1\bar{e}& \ &\text{SI} \\
		                                                 &\S^{(m+2)}+ \ 2\bar{e}& \  &\text{DI} \\
		                                                &\S^{(m+3)}+ \ 3\bar{e}& \ &\text{TI}
		  \end{align*}

We consider the case of the photoionisation of the \S$^{+}$ ion in the first instance. If the energy of the incident photon is below the 2p or 2s threshold energies, then $2\text{p} \rightarrow n\text{d} \ \text{,}\ (n+1)\text{s}$ and $2\text{s} \rightarrow n\text{p}$ excitations will respectively produce Rydberg series of  $2p^53s^23p^3n\text{d}, 2p^53s^23p^3(n+1)\text{s\, and} \  2s2p^63s^23p^3n\text{p}$ autoionising states appearing as well-developed resonances in the cross section spectrum of the photoabsorption continuum. Of the first two series, the $2\text{p}\rightarrow n\text{d}$ oscillator strength is expected to be at least nine times stronger than the $2\text{p}\rightarrow n\text{s}$ strength due to angular momentum coupling, or geometrical factors ~(see, e.g.,~\cite{friedrich}). The latter absorption gives rise to $2\text{s}n\text{p}(^4\text{S}, ^4\!\text{P})$ resonance series that converge to the higher-energy $2\text{s}^{-1}$ L-vacancy state.The specific inner-shell photoexcitation processes to be considered here are summarised, from a single-configuration, non-relativistic (LS-coupled) perspective, in Table~\ref{Excitation Schemes} for each of the \S$^{+}$, \S$^{2+}$ and \S$^{3+}$ ions.  Each intermediate autoionizing, or resonant,  $2p^53s^23p^3n\text{d}$ state can decay via either of the participator or the spectator resonant Auger processes.  In participator Auger decay, e.g. the $L_{2,3}\!-\!M_1M_{4,5}$,  $2p^53s^23p^3n\text{d} \rightarrow  2p^63s3p^3 +\bar e$, Auger transition the valence electron $n\text{d}$ participates in the decay, giving a rate that scales as $1/n^3$. Spectator Auger decay, e.g. the $L_{2,3}\!-\!M_1M_{1}, \,  2p^53s^23p^3nd  \rightarrow  2p^63p^3nd +\bar e$, Auger transition, 
 proceeds via a stronger, $n$-independent Auger rate, broadening the entire Rydberg series of resonances below the L-edge. All the possible spectator and participator $\text{L}_{2,3}\!-\!{\text{M}}_x{\text{M}}_y$ and $\text{L}_{1}\!-\!{\text{M}}_x{\text{M}}_y (\ x\leq y=1\ {\text{or}}\ 2,3)$ Auger decay transitions from  $2p^53s^23p^33\text{d}$ and $2s2p^63s^23p^4$ in \S$^{+}$, $2p^53s^23p^23\text{d}$ and $2s2p^63s^23p^3$ in \S$^{2+}$,\  and $2p^53s^23p3\text{d}$ and $2s2p^63s^23p^2$ in \S$^{3+}$, respectively, are shown in Table~\ref{Excitation Schemes}. From an experimental viewpoint, the various non-radiative decay schemes just described are the main contributors to the production of ions in the single-ionisation channel (SI).
\begin{turnpage}
\begingroup
\squeezetable
\begin{table*} 
\caption{\label{Excitation Schemes}2p and 2s resonant excitations (strict LS selection rules) and subsequent Auger decay routes, metastable states, successive ionization energies and inner thresholds in the \S$^{+}$, \S$^{2+}$ and \S$^{3+}$ isonuclear sequence}
\begin{ruledtabular}
          \begin{tabular}{rlll} 
          \multicolumn{1}{c}{} & \multicolumn{1}{c}{\textrm{\S$^+$ (phosphor-like)}} & \multicolumn{1}{c}{\textrm{\S$^{2+}$ (silicon-like)}} &\multicolumn{1}{c}{\textrm{\S$^{3+}$ (aluminium-like)}} \\
	 \cline{2-2} \cline{3-3} \cline{4-4}
	 \vspace{-0.15cm}\\
	 Ground State GS (eV)  & $2p^6 3s^2 3p^3 \ {^4}\text{S}_ {\!  {\frac{3}{2}}} \  (0.00)$ & $2p^6 3s^2 3p^2 \ {^3}{\text{P}}_{\! 0} \ (0.00)$  &  $2p^6 3s^2 3p \ {^2}{\text{P}}_{\! \frac{1}{2}}  \ (0.00)$\\
	 \vspace{-0.15cm}\\
	 First metastable states (eV)\footnotemark[1] & $2p^6 3s^2 3p^3 \ {^2}\text{D}  \ (1.84),\ {^2}\text{P} \ (3.05) $ & $2p^6 3s^2 3p^2 \ {^1}{\text{D}}_{\! 2} \ (1.40), \ {^1}{\text{S}}_{\! 0} \ (3.37) $ & $2p^6 3s3p^2 \ {^4}{\text{P}}_{\! \frac{1}{2}} \ (8.83)\footnotemark[2] $  \\
	 \vspace{-0.15cm}\\
	 Single, double and triple ionisation energies (eV)\footnotemark[1] & 23.34, 58.20, 105.42 & 34.86, 82.08, 154.67 & 47.22, 119.81, 207.86 \\
	\vspace{-0.15cm}\\
	 $2p \rightarrow nd$ excitations\footnotemark[3] \footnotemark[4](n$\geqslant 3)$ & $2p^5\big(3s^23p^3n\text{d}\ ^3\text{D}\big)\  \ ^{4}\text{P} $ & $2p^5\big(3s^23p^2nd\ ^2\text{P}\big)\  \ ^{3}\text{S}\ ^{3}\text{P} \  ^{3}\text{D}$ & $2p^5\big(3s^23pnd\ ^1\text{P}\big)\ ^{2}\text{S}\  ^{2}\text{P}\  ^{2}\text{D}$\\
	 & $2p^5\big(3s^23p^3n\text{d}\ ^5\text{D}\big)\  \ ^{4}\text{P} $ & $2p^5\big(3s^23p^2nd\ ^4\text{P}\big)\  \ ^{3}\text{S}\ ^{3}\text{P} \ ^{3}\text{D}$ & $2p^5\big(3s^23pnd\ ^3\text{P}\big)\ ^{2}\text{S}\  ^{2}\text{P}\  ^{2}\text{D}$\\
	  &                                                                                                                          &  $2p^5\big(3s^23p^2nd\ ^2\text{D}\big)\  \ ^{3}\text{P}\ ^{3}\text{D} $                                                                    &   $2p^5\big(3s^23pnd\ ^1\text{D}\big)\ ^{2}\text{P}\  ^{2}\text{D}$ \\
	  &                                                                                                                          &  $2p^5\big(3s^23p^2nd\ ^4\text{D}\big)\  \ ^{3}\text{P}\ ^{3}\text{D} $                                                                    &   $2p^5\big(3s^23pnd\ ^3\text{D}\big)\ ^{2}\text{P}\  ^{2}\text{D}$ \\
	 \vspace{-0.15cm}\\
	 2p $\rightarrow (n+1)s $ excitations\footnotemark[4] $(n\geqslant 3)$ & $2p^5\big(3s^23p^3(n+1)\text{s}\ ^3\text{S}\big)\ ^{2,4}\text{P}$ & $2p^5\big(3s^23p^2(n+1)s\ ^2P\big) \ ^{1\!,3}\text{S} \  ^{1\!,3}\text{P} \ ^{1\!,3}\text{D} $& $2p^5\big(3s^23p(n+1)s\ ^1\text{P}\big) \ ^{2}\text{S}\ ^{2}\text{P}\ ^{2}\text{D}$\\
	  & $2p^5\big(3s^23p^3(n+1)\text{s}\ ^5\text{S}\big)\ ^{4,6}\text{P}$&$2p^5\big(3s^23p^2(n+1)s\ ^4P\big) \ ^{3\!,5\!}\text{S} \  ^{3\!,5\!}\text{P} \ ^{3\!,5\!}\text{D} $ & $2p^5\big(3s^23p(n+1)s\ ^3\text{P}\big) \ ^{2,4}\text{S}\ ^{2,4}\text{P}\ ^{2,4}\text{D}$\\
	  \vspace{-0.15cm}\\
	 2s $\rightarrow n$p excitations\footnotemark[4]  & $2s2p^63s^2\big(3p^3n\text{p}\ ^3\text{P}\big)\ ^{4}\text{P}$ & $2s2p^63s^23p^2n\text{p}\ ^{3}\text{S}\ ^{3}\text{P}\ ^{3}\text{D}$ & $2s2p^63s^23pn\text{p}\ ^{2}\text{S}\ ^{2}\text{P}\ ^{2}\text{D}$\\
	 & $2s2p^63s^2\big(3p^3n\text{p}\ ^5\text{P}\big)\ ^{4}\text{P}$ &  & \\
	 \vspace{-0.15cm}\\
	 $2p$ inner-thresholds & $2p^53s^23p^3\ ^{3,5}\text{P}$ & $2p^53s^23p^2(^3\!\text{P}) ^{2\!,4\!}\text{S} \ ^{2\!,4\!}\text{P} \ ^{2\!,4\!}\text{D} $& $2p^53s^23p\ ^{1,3}\text{S}\ ^{1,3}\text{P}\ ^{1,3}\text{D}$ \\
	 \vspace{-0.15cm}\\
	$2s$ inner-thresholds & $2s2p^63s^23p^3\ ^{3,5}\text{S}$ & $2s2p^63s^23p^2\ ^{2,4}\text{P}$ & $2s2p^63s^23p\ ^{1,3}\text{P}$\\
	\vspace{-0.15cm}\\
	 $L_{2,3}\!-\!M_xM_{y}$\ Auger transitions\footnotemark[5] & $2p^63s^23p^2,3s3p^3,3s3p^23d,$& $2p^63p^23d,3s3p3d,3s3p^2,$ & $2p^63p3d, 2p^63s3d, 2p^63s3p$\\
	 & $3p^33d,3s^23p3d$ & $2p^63s^23d, 3s^23p$ & $2p^63s^2$  \\
	 \vspace{-0.15cm}\\
	 $L_{1}\!-\!M_xM_y,\ L_{1}\!-\!L_{2,3}M_y$ \ Auger transitions\footnotemark[6]  & $2s^22p^53s^23p^3,3s3p^4$& $2s^22p^53s^23p^2, 2p^53s3p^3$ & $2s^22p^53s3p^2, 2p^53s^23p$\\
	 & $2s^22p^63p^4,3s3p^3,3s^23p^2$ & $2s^22p^63s3p^2, 2p^63p^3 $  & $2s^22p^63p^2, 2p^63s3p$\\
	 & $2s^22p^63s^23p^2$ & $2s^22p^63s^23p$& $2s^22p^63s^2$\\
	 \end{tabular}
\end{ruledtabular}
\footnotetext[1]{See Ref.~\citep{Kramida_2023} for more details.}
\footnotetext[2]{This state is not metastable}
\footnotetext[3]{Not all the possible final LS states are shown}
\footnotetext[4]{From GS only}
\footnotetext[5]{Initial configuration $2p^53s^23p^33\text{d}$ (\S$^{+}$), $2p^53s^23p^23\text{d}$ (\S$^{2+}$), $2p^53s^23p3\text{d}$ (\S$^{3+}$).}
\footnotetext[6]{Initial configuration $2s2p^63s^23p^4$ (\S$^{+}$), $2s2p^63s^23p^3$ (\S$^{2+}$), $2s2p^63s^23p^2$ (\S$^{3+}$).}
\end{table*}
\endgroup
\end{turnpage}

If the photon energy is greater than the 2p ionisation threshold energy (see Table \ref{Excitation Schemes}), then a vacancy is created in the 2p-subshell which can readily decay via Auger emission. This atomic decay process appears as a continuum and the dominant one in the double-ionisation channel (DI). The initial 2p vacancy may also decay via the double-Auger emission process leading to a continuum signal in the triple-ionisation channel (TI) which is comparatively quite weak. Discrete resonances in the DI and TI channels are typically linked with higher order electron correlation effects present during the single electron Auger emission process for a given resonance, and thus observed at the same photon energies as in the SI channel. These effects, eg. shake-off, result in the loss of one or two additional electrons.

	\subsection{\label{mcdf}Multiconfigurational Dirac-Fock Calculations}
The MCDF calculations were performed with the help of an updated version of the code originally developed by Bruneau \citep{Bruneau_1984}, using a full intermediate coupling scheme based on a $jj$ basis set. Photo-excitation and direct photo-ionization cross sections were computed in the Babushkin gauge \citep{Grant_1974} considering electric dipole transitions only. A review of state-of-the-art developments in relativistic multiconfiguration atomic structure calculations can be found in \citep{MCDF_2023}. For each of the \S$^{+}$, \S$^{2+}$ and \S$^{3+}$ ion stages, a preliminary computational step consisted of optimizing a unique set of bound radial functions needed for the overall photo-absorption processes under consideration. For \S$^{+}$, the orbital set optimization was deduced from the minimization of a pseudo-state energy involving two subsets of configurations. The first subset (labelled $L$) brings together the configurations: [Ne]$3s^23p^3$, [Ne]$3s3p^33d$, [Ne]$3p^33d^2$, and [Ne]$3s^23p^2nl$, while the second (labelled $U$) consists of the configurations: [Ne]$3s^23p^2$, [F]$3s^23p^3$, and [F$^*$]$3s^23p^3$ (where [Ne]$=1s^22s^22p^6$, [F]$=1s^22s^22p^5$, [F*]$=1s^22s2p^6$,  $n=4,...,7$ and $l=s,p,d$). The pseudo-state energy is defined as if $J$ is a total spin-orbital quantum number. For \S$^{2+}$, the $L$-subset brings together the configurations: [Ne]$3s^23p^2$, [Ne]$3s3p^23d$, [Ne]$3p^23d^2$, [Ne]$3p^24s^2$, and [Ne]$3s^23pnl$, and the $U$-subset brings together the configurations: [Ne]$3s^23p$, [F]$3s^23p^2$, and [F*]$3s^23p^2$. For \S$^{3+}$, the $L$-subset consists of the [Ne]$3s^23p$ and [Ne]$3s^2nl$ configurations, and the $U$-subset brings together the [Ne]$3s^2$, [F]$3s^23p$, and [F*]$3s^23p$ configurations. 

Once a specific set of bound radial functions was obtained, the photo-excitation cross sections from the $2s$ and $2p$ subshells were separately computed, as well as the direct $2p$, $3s$, and $3p$ photo-ionization cross sections. These calculations were performed for all the levels belonging to the ground configuration of each ion stage. For \S$^{+}$, the $^4S_{3/2}, \ ^2D_{3/2,5/2}\ \text{and} \ ^2P_{1/2,3/2}$ levels were considered. To describe these initial states, a multiconfiguration expansion was used involving the configurations [Ne]$3s^23p^3$, [Ne]$3s3p^33d$, and [Ne]$3p^33d^2$. To describe the $2p$ photo-excited levels, the [F]$3s^23p^33d$, [F]$3s^23p^3ns$, and [F]$3s^23p^3nd$ configurations were selected while the configurations [F*]$3s^23p^4$ and [F*]$3s^23p^3np$ were considered to describe the excited levels resulting from the photo-excitation of the $2s$ subshell. The $2p$ photo-ionization cross sections were calculated using the [Ne]$3s^23p^3$ and [F]$3s^23p^3$ configurations, while the $3s$ and $3p$ photo-ionization cross sections were calculated using the [Ne]$3s^23p^3$, [Ne]$3s^23p^2$, and [Ne]$3s3p^3$ configurations. 
Similar schemes were used for \S$^{2+}$ and \S$^{3+}$. For \S$^{2+}$, the $^3P_{0,1,2},\ ^1D_2\  \text{and}\  ^1S_0$ initial levels were considered and the [Ne]$3s^23p^2$, [Ne]$3s3p^23d$, [Ne]$3p^23d^2$ and [Ne]$3p^24s^2$ configurations retained to describe them. The last three configurations were found to have the largest effect on the energies of the $^3P_{0,1,2},\ ^1D_2\  \text{and}\  ^1S_0$ states, due to configuration mixing effects.
The [F]$3s^23p^23d$, [F]$3s^23p^2ns$ and [F]$3s^23p^2nd$ configuration expansion was used to describe the excited levels resulting from the $2p$ photo-excitation while  [F*]$3s^23p^3$ and [F*]$3s^23p^2np$ were selected to describe the excited levels resulting from the photo-excitation of the $2s$ subshell. The [Ne]$3s^23p^2$ and [F]$3s^23p^2$ configurations were used to evaluate the $2p$ photo-ionization cross sections while the $3s$ and $3p$ photo-ionization cross sections were calculated using the [Ne]$3s^23p^2$, [Ne]$3s^23p$ and [Ne]$3s3p^2$ configurations. For \S$^{3+}$, the [Ne]$3s^23p$, [Ne]$3s3p3d$, [Ne]$3p3d^2$, and [Ne]$3p^3$ configurations were retained to describe the $^2P_{1/2,3/2}$ initial levels. The [F]$3s^23p3d$, [F]$3s^23pns$ and [F]$3s^23pnd$ configuration expansion was used to describe the $2p$ photo-excited excited levels, while the [F*]$3s^23p^2$, [F*]$3s^23pnp$ were selected to describe the excited levels resulting from the photo- excitation of the $2s$ subshell. The [Ne]$3s^23p$ and [F]$3s^23p$ configurations were used to evaluate the $2p$ photo-ionization cross sections while the $3s$ and $3p$ photo-ionization cross sections were calculated using the [Ne]$3s^23p$, [Ne]$3s^2$, and [Ne]$3s3p$ configurations.

\subsection{Overview of the R-matrix Method (Breit-Pauli and Dirac Formulations)}

\begin{figure}
\includegraphics[width=8.6cm]{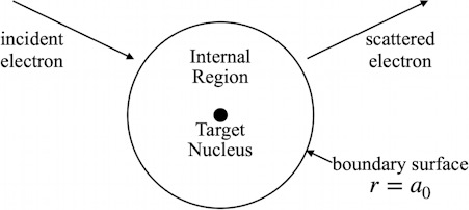}
\caption{\label{figinnerouter}Separation of Configuration Space in the R-matrix Method}
\end{figure}

We use two completely independent R-matrix~\citep{burke} suites of codes, one based on a Breit-Pauli formalism~\citep{nigelrmat} and the other a Dirac formalism~\citep{connorrmat,Grant_2006}.  In either case, configuration space for the outer-most electron is divided into inner, multi-electron interacting region and outer, asymptotic region (see figure ~\ref{figinnerouter}) with the matching of inner and outer region wavefunctions achieved via use of the R-matrix, as defined below.

In either of these computer packages, the essential expansion of the wavefunction of the final state as Rydberg-like quasibound states embedded in one or more continua takes on a close coupling prescription~\citep{seatoncc}.

  The total wave function is expanded for each spatial symmetry as
\begin{eqnarray*}
    \Psi & = & \sum_{ij} c_{ij} \Phi_i u_{j} + \sum_{k} b_{k} \chi_k\ ,
    \end{eqnarray*}
where $\Phi_i$ are the target ionic states, the electron continuum orbitals are spanned by the basis $u_{ij}$, and the short-range correlation polarization is linked to additional basis functions $\chi_k$.

In the outer region, the total wave function is expanded in terms of spherical harmonics $Y_{l_jm_j}$:
\begin{eqnarray*}
    \Psi_k(\bf{x_1, x_2...x_{N+1}}) = & \sum_i \sum_{j} \frac{F_j(r_{N+1})}{r_{N+1}}Y_{l_jm_j} \Phi(\bf{x_1, x_2...x_{N}})\ .
\end{eqnarray*}
Here the $\bf{x_i}$ denote the spatial and spin variables of the $i^{th}$-electron, with implied spin coupling of the $(N+1)^{th}$-electron's spin to the target functions $\Phi(\bf{x_1, x_2...x_{N}})$.

A matching condition at the boundary $r=a_0$ is facilitated by the inverse-log-like "R-matrix" {\bf R}, defined as
\begin{eqnarray*}
{\bf R} & \equiv & \frac{1}{a_0} ({\bf F}^\prime)^{-1} {\bf F}\ ,
\end{eqnarray*}
where {\bf F} and {\bf F}$^\prime$ are the inner-region function and derivative matrices for the outer electron.  The power of the inner-region basis expansion is to derive an analytical expression for the R-matrix, at all energies $E$, in terms of eigenvalues and eigenvectors of the inner-region Hamiltonian and their basis-function surface amplitudes,
\begin{eqnarray*}
{\bf R}(E) & = & \frac{1}{a_0}{\bf w}\,\left({\bf e} - E\right)^{-1}{\bf w}^T \ ,
\end{eqnarray*}
where {\bf e} are the eigenvalues of the inner-region Hamiltonian and {\bf w} represents a matrix of eigenvectors projected onto the surface $r=a_0$.

Given the outer-region asymptotic channel form
\begin{eqnarray*}
{\bf F^{outer}} & = & {\bf f}^- - {\bf f}^+\,{\bf S}\ ,
\end{eqnarray*}
in terms of incoming and outgoing, long-range Coulomb functions  {\bf f}$^{\pm}$, the scattering matrix ${\bf S}$ can be determined, thereby yield separate wavefunctions for the initial bound state $\Psi_i$ and the final structured-continuum state $\Psi_f$.  The latter, by design, include all quasi-bound resonance states, and all open continuum states, all coupled and therefore correlated acccording to the convergence reached with the given basis.  The dipole matrices $D$ can be numerically evaluated to compute photoabsorption (photoexcitation and photoionization) cross sections:
\begin{eqnarray*}
\sigma_{PA} & = & \frac{4\pi\alpha}{3}\frac{g_f}{g_i}\vert \langle
\Psi_i \vert D\vert\ \Psi_f\rangle\vert^2\ ,\label{eqtotal2}
\end{eqnarray*}
where $g_i$ and $g_f$ are the statistical weights of the initial and final states, respectively, and $\alpha$ is the fine-structure constant.
The above procedure is now applied to the specific sulphur ions treated here using either the BPRM or DARC formalisms.

	 \subsection{\label{rmatrix} Breit-Pauli R-Matrix Calculations}
To compute the L-shell photoionization of the S$^{3+}$ and S$^{2+}$ ionic states, the BPRM method was used ~\cite{burke,nigelrmat}. In order to include the important configuration interaction effects anticipated, the initial and final ionic states were constructed from linear combinations of coupled configurations of atomic orbitals spanned by a basis similar in principle to earlier Mg-like S$^{4+}$ calculations ~\citep{Mosnier_2022}, or by even more recent BPRM calculations for neutral sulphur~\citep{Gatuzz_2023}, neutral sodium~\citep{gorczyca2024}, neutral argon~\citep{Gatuzz_2024}, and ionized chlorine~\citep{Mosnier_2023}.  This atomic structure expansion is large enough to account for anticipated higher-order electronic promotions from initial states while still being
computationally feasible.

Beginning with the S$^{3+}$ ground state within a single configuration perspective in a non-relativistic, LS- coupling scheme, the specific processes to be considered are given as
\begin{eqnarray*}
\label{eq1}
h\nu+2s^22p^63s^23p(^2P) & \rightarrow &
2s^22p^53s^23pnd(^{2\!}S,^{2\!\!}P,^{2\!\!}D) \ , \\
& & 2s^22p^53s^23pns(^{2\!}S,^{2\!\!}P,^{2\!\!}D) \ , \\
& & 2s2p^63s^23pnp(^{2\!}S,^{2\!\!}P,^{2\!\!}D) \ .
\end{eqnarray*}
These processes give rise to nine dominant Rydberg series of absorption lines, or resonances, in the resultant photoionization cross section.  The stronger $2p\rightarrow nd$ and weaker $2p\rightarrow ns$ series each converge to one of the various $2p^53p(^{1,3}(SPD))$ S$^{4+*}$ excited thresholds whereas at higher energies the two $2s\rightarrow np$ series converge to the $2s3p(^3P)$ and $2s3p(^1P)$ thresholds.

These photoexcited intermediate states, such as the $2p^53s^23pnd$ resonances, can then, as previously noted, autoionize to the S$^{4+}$ continua via one of
two qualitatively different decay pathways.  First there is {\em participator} Auger decay
\begin{eqnarray}
 2p^53s^23pnd & \rightarrow & 2p^63s^{(1+a)}3p^b +\bar{e} \ \ \ (a+b = 1)\nonumber \ ,
\label{eqpart3+}
\end{eqnarray}
for which the valence electron $nd$ {\em participates} in the autoionization process, therefore giving an Auger width that scales as $1/n^3$.
The second pathway is {\em spectator} Auger decay
\begin{eqnarray}
 2p^53s^23pnd & \rightarrow & 2p^63s^a3p^bnd +\bar{e}  \ \ \ (a+b = 1)\nonumber \ ,
\label{eqspect1}
\end{eqnarray}
for which the width is independent of $n$ and therefore 
broadens the Rydberg series near threshold.

Participator Auger decay process was explicitly implemented in a straightforward manner with the inclusion of the $2p^63s^{(1+a)}3p^b \ (a+b = 1)$ decay channels in the standard R-matrix theory \citep{burke,berrington95}.  On the other hand, spectator Auger decay was implicitly treated via an optical potential approach~\citep{Gorczyca99} that introduces an imaginary potential $-i \Gamma/2$ to the multichannel quantum defect scattering formulation, where $\Gamma$ is the spectator Auger width that can be computed separately or derived empirically from experimental results.  Here it was found that the Lorentzian Auger width is much less than the broader Gaussian width due to experimental broadening, so a fixed spectator Auger width $\Gamma=2\times 10^{-4}\, {\rm Ryd}=2.7$~meV was chosen that is much less than the Gaussian broadening of $\approx$ 160~meV but larger than the energy mesh step of 1.4~meV used in the final R-matrix calculation.

For photoionization of the S$^{2+}$ ground state, again depicted from a single configuration, LS-coupled perspective, the following processes are relevant:
\begin{eqnarray*}
\label{eq2}
h\nu+2s^22p^63s^23p^2(^{3\!}P) & \rightarrow &
2s^22p^53s^23p^2nd(^{3\!}S,^{3\!\!}P,^{3\!\!}D) \ , \\
& & 2s^22p^53s^23p^2ns(^{3\!}S,^{3\!\!}P,^{3\!\!}D) \ , \\
& & 2s2p^63s^23p^2np(^{3\!}S,^{3\!\!}P,^{3\!\!}D) \ .
\end{eqnarray*}
These absorption lines are properly included in the R-matrix theory to benchmark prominent resonances in the experimental photoionization cross section.   
The subsequent {\em participator} Auger decay processes are given by
\begin{eqnarray}
 2p^53s^23p^2nd & \rightarrow & 2p^63s^{(1+a)}3p^{(1+b)} +\bar{e} \ \ \ (a+b = 1)\nonumber \ ,
\label{eqautoion}
\end{eqnarray}
Whereas the {\em spectator} Auger decay channels are
\begin{eqnarray}
 2p^53s^23p^2nd & \rightarrow & 2p^63s^{(1+a)}3p^bnd +\bar{e}  \ \ \ (a+b = 1)\nonumber \ .
\label{eqspect2}
\end{eqnarray}

The complete wavefunctions of the initial bound, intermediate quasibound, and final continuum states were constructed from coupled configurations of both physical atomic orbitals ($1s$, $2s$, $2p$, $3s$, $3p$, $3d$) and additional pseudoorbitals 
($\overline{4s}$, $\overline{4p}$, $\overline{4d}$) optimized to account for orbital relaxation effects caused by the $2s$ or $2p$ vacancies.  More specifically, for the S$^{3+}$ photoionization calculations, the $1s$, $2s$, $2p$, and $3s$ physical orbitals were obtained from a single-configuration Hartree-Fock (HF) calculations on the S$^{4+} (1s^22s^22p^63s^2)$ ground-states. Additional {\em physical} $3p$ and $3d$ orbitals were generated from frozen-core HF calculations on the S$^{4+}$
$(1s^22s^22p^63s3p)$ and $(1s^22s^22p^63s3d)$ excited states.  Additional $\overline{4s}$, $\overline{4p}$,  and $\overline{4d}$ {\em pseudoorbitals } were generated from a multi-configuration Hartree-Fock (MCHF) calculation on the inner-shell-vacancy $2p^53s3p3d$ ionized state of S$^{4+}$ including single and double promotions from this configuration to other physical orbitals and pseudoorbitals (e.g., $2p\rightarrow {3p}$ and $2p\rightarrow \overline{4p}$).

Similarly, for the S$^{2+}$ photoionization calculations, $1s$, $2s$, $2p$, $3s$, and $3p$ physical orbitals were obtained from a single-configuration Hartree-Fock (HF) calculation on the S$^{3+}$ $(1s^22s^22p^63s^23p)$ ground-state. An additional $3d$ {\em physical} orbital was generated from a frozen-core HF calculation on the S$^{3+}$ $(1s^22s^22p^63s^23p3d)$ excited states. The $\overline{4s}$, $\overline{4p}$,  and $\overline{4d}$ {\em pseudoorbitals } were generated from a multi-configuration Hartree-Fock (MCHF) calculation on the inner-shell-vacancy $2p^53s^23p3d$ ionized state of S$^{3+}$ including single and double promotions from this configuration.
An additional orbital basis of 60 R-matrix-generated continuum orbitals were then coupled to configurations of physical and pseudo atomic orbitals to make up the basis of configurations used to span the total wavefunctions of the initial bound, intermediate quasibound, and final continuum states.
Lastly, for computing photoionization cross sections from the metastable $S^{3+}(2p^63s3p^2(^4P))$ and $S^{2+}(2p^63s^23p^2(^1D))$ states of $S^{3+}$ and $S^{2+}$, respectively, the same procedure was used as for the ground states.

The methodology detailed above for the 
 S$^{3+}$ and S$^{2+}$ ions was also attempted for the S$^{+}$ ion.  However, because the complexity of the problem grew as the ionic charge was reduced (as more electrons were added), the same orbital basis approach, with the increased number of configurations needed for the S$^+$ ion, led to a problem that was too computationally demanding on the Linux workstations available to us at the moment.  Fortunately, earlier unpublished work using the Dirac R-matrix approach had already been carried out on a larger massively parallel machine, and those calculations and results are described in the next subsection.

\subsection{\label{darc}Dirac $R$-matrix calculations}
To perform L-shell photo-ionization of S$^+$ which incorporates a large number of coupled states and channels, it was necessary to utilize the Dirac Atomic $R$-matrix (DARC) relativistic parallel codes designed to run efficiently on Massive Parallel Architectures ~\citep{connorrmat}. Two types of DARC models were employed, for the residual S$^{2+}$ target ion. The 
first utilized an $n = 3$ orbital basis set and the second a more elaborate model that used an 
$n = 4$ orbital basis set. In both models 12 continuum orbitals were used for the $n = 3$
and the more diffuse $n = 4$ basis.

Model A used an $n=3$ orbital basis, with 5 configurations included from the valence states: 1s$^2$2s$^2$2p$^6$3s$^2$3p$^2$, 3s3p$^3$, 3s$^2$3p3d, 3s$^2$3d$^2$, 3s3p$^2$3d and 4 configurations from the 2p$^{-1}$ hole 
states : 1s$^2$2s$^2$2p$^5$3s$^2$3p$^3$, 2p$^5$3s3p$^4$, 2p$^5$3s$^2$3p$^2$3d and 
2p$^5$3s3p$^3$3d. These give rise to a total 726 coupled states. In order to fit the
photoionization dynamical calculations into the computer our target was 
reduced to the lowest 579  coupled states.

Model B used an $n=4$ orbital basis with 10 configurations from the valence states : 1s$^2$2s$^2$2p$^6$3s$^2$3p$^2$,
3s3p$^3$, 3s$^2$3p3d, 3s$^2$3p4s, 3s$^2$3p4p, 3s$^2$3p4d, 3s$^2$3d$^2$, 3s$^2$4s$^2$,
3s$^2$4p$^2$, 3s$^2$4d$^2$  and 6 configurations from the 2p$^{-1}$ hole states:
1s$^2$2s$^2$2p$^5$3s$^2$3p$^3$, 2p$^{5}$3s3p$^4$, 2p$^5$3s$^2$3p$^2$3d. This gave rise to 619 target states. Here again in order to fit the photoionization  dynamical calculations 
into the computer our target was reduced to the lowest 527 coupled states.

\begin{figure}
\includegraphics[width=86mm]{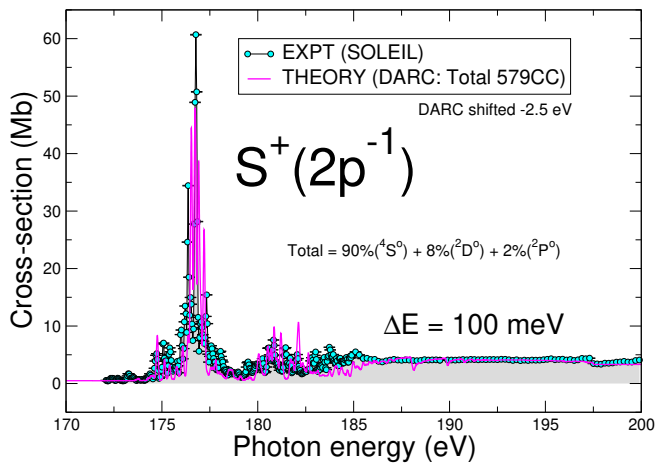}
\caption{\label{DARC}Comparison of the DARC results from model A with the experimental
          results from SOLEIL taken at a band pass of 100meV.  The 
          theoretical results have been convolved with a normalised Gaussian function
          having a width of 100 meV and are
          given for a beam mixture of 
          90\%($^4$S$^{\text{o}}$) + 8\%($^2$D$^{\text{o}}$) + 2\%($^2$P$^{\text{o}}$).}
\end{figure}

We performed photoionization cross-section calculations 
in both DARC models for photoionization out of the 
$^4$S$^{\text{o}}_{3/2}$ ground state and the metastable states $^2$D$^{\text{o}}_{5/2,3/2}$ and $^2$P$^{\text{o}}_{3/2,1/2}$. In our calculations to fully 
resolve resonance features in the appropriate 
photoionization cross section the outer region electron ion collision complex was solved 
using a fine energy mesh of 1.36 meV. From a direct comparison of the DARC theoretical cross sections  
with the experimental photoionization data, we estimated a best fit beam mixture of 90\% $^4$S$^{\text{o}}_{3/2}$, 8\% $^2$D$^{\text{o}}$ and 2\% $^2$P$^{\text{o}}$.   
Model A gave slightly better bound state 
energies and cross sections for S$^+$ than model B 
for the metastable states. The results from Model A with 579 coupled states are plotted in figure~\ref{DARC}, 
demonstrating the fairly good agreement achieved by including the large number of levels. The bound states energies of S$^+$ are compared with NIST~\cite{Kramida_2023}experimental energies in Table~\ref{DARCTable}. Model A generally gives a better energy agreement as it includes all the one- and two-electron promotions whereas Model B contains all the one-electron promotions and a limited number of two-electron promotions only.

\begin{table}
\begin{ruledtabular}
\begin{tabular}{cccccc}
S$^+$   State   &DARC    &DARC      &NIST       &       &       \\
                & $n=3$     &$n=4$       &           &       &       \\
                & Model A    & Model B &           &       &       \\
                &$E$ (Ry)  &$E$ (Ry)      &$E$ (Ry)      &$\Delta_1$&$\Delta_2$\\
 \hline               
$^4$S$^{\text{o}}_{3/2}$ &1.85829 &1.78234   &1.71874    & 8.0   &3.7   \\ 
$^2$D$^{\text{o}}_{3/2}$ &1.71743 &1.66384   &1.58339    & 8.5   &5.6   \\
$^2$D$^{\text{o}}_{5/2}$ &1.71361 &1.78202   &1.58310    & 8.3   &12.6  \\
$^2$D$^{\text{o}}_{1/2}$ &1.62598 &1.66445   &1.49525    & 8.7   &11.3  \\
$^2$D$^{\text{o}}_{3/2}$ &1.61868 &1.66387   &1.49483    & 8.3   &11.3  \\
\end{tabular}
\end{ruledtabular}
\caption{\label{DARCTable}Bound states of S$^+$ compared with experimental energies from
        NIST\cite{Kramida_2023}. $\Delta_1$ is the percentage difference for Model A and $\Delta_2$ that for Model B. All energies are in Rydbergs.}
\end{table}

\section{\label{resultsandanalyses}Results and Analyses}
\begin{figure}
         \includegraphics[width=86mm]{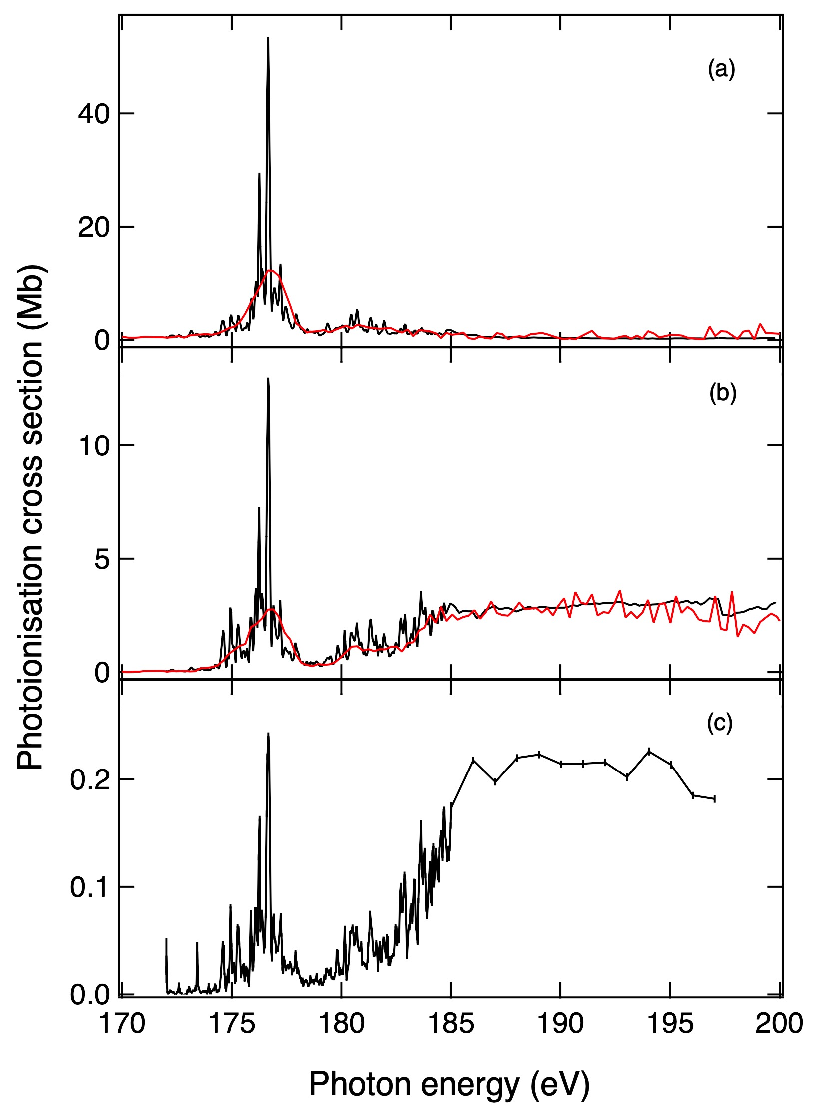}
          \caption{\label{S+exp}Absolute photoionisation cross sections of singly ionised (\S$^+$) sulphur measured, with a photon energy bandpass of $\approx$ 100 meV, in the (a) single  ionisation 	(SI), (b) double ionisation (DI) and (c) triple ionisation (TI) channels, respectively. The SI and DI red traces are cross section values obtained at the ASTRID synchrotron by Kristensen \textit{et al.} 	\citep{Kristensen_2002}}
         \end{figure}

	 \begin{figure}
	\includegraphics[width=86mm]{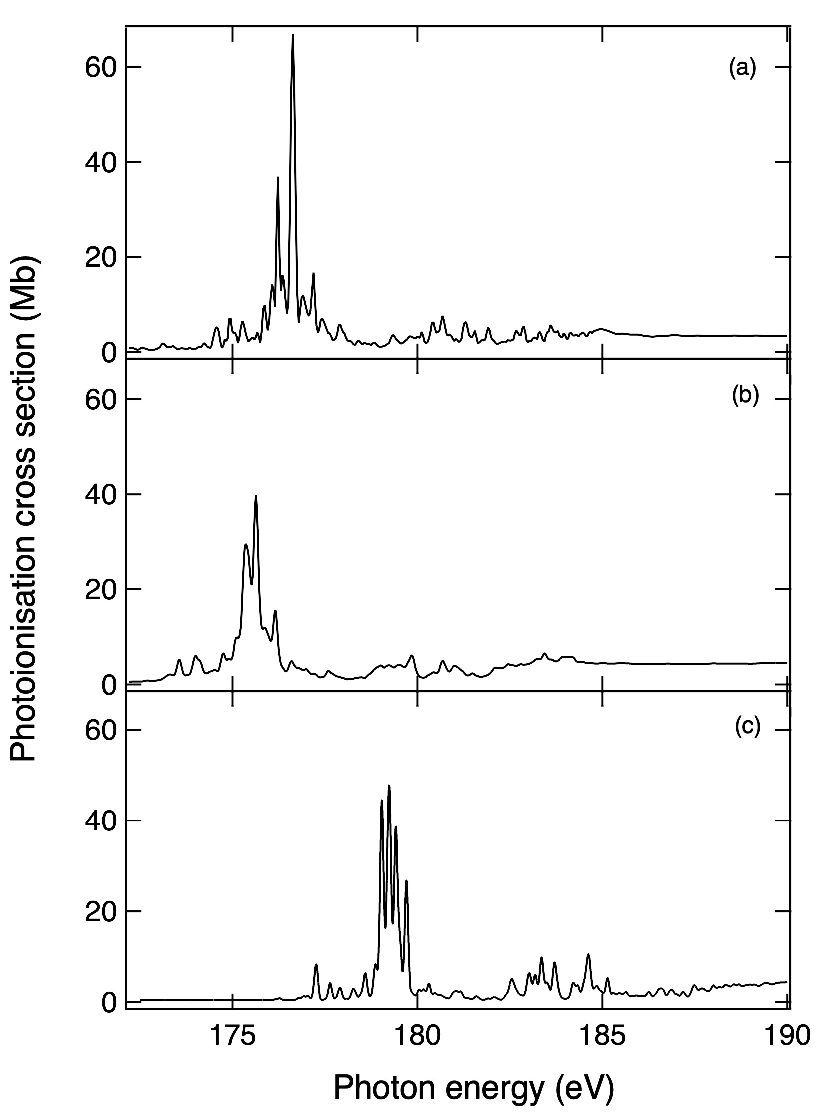}
	\caption{\label{S+exptheo}Total photoionisation cross sections of singly ionised sulphur \S$^+$ in the region of the 2p threshold: (a) Experimental values obtained in this work, (b) 	multiconfigurational Dirac-Fock (MCDF) theoretical theoretical values, (c) Dirac Atomic R-Matrix Codes (DARC) theoretical values. The MCDF and DARC theoretical cross sections were convolved with a normalised Gaussian function of FWHM 100 meV to simulate the experimental broadening of FIG.~\ref{S+exptheo}(a). The  MCDF and R-Matrix theoretical intensities were scaled to account for metastable state populations (see text).}
 	\end{figure}

	\subsection{\label{P-like}Phosphorus sequence: S$^+$ }
	The measured SI, DI and TI cross sections for \S$^{+}$ are presented in Figures~\ref{S+exp} (a),(b) and (c), respectively. Summing the SI, DI and TI results yields the total experimental photoionisation cross sections. These are compared as a function of photon energy, with the energy-unshifted MCDF and DARC R-Matrix theoretical results,  in Figures~\ref{S+exptheo} (a),(b) and (c), respectively. From the integration of the SI, DI and TI cross section curves in the 170-180 eV photon energy range, we obtain intensity ratios of $\text{DI/SI}= 28\%$ and $\text{TI/SI}= 0.07\%$. Based on the MCDF results, we can assign the multiple intense resonances observed in the 173-181 eV photon range to the autoionizing decay of a number of the many possible atomic states built on the $2p^53s^23p^33\text{d}$ configuration. Unconstrained least-square fitting of the data of Fig.~\ref{S+exptheo}(a) using Gaussian functions of full width at half maximum (FWHM) 150 meV, provides the resonance energy positions (in eV) and strength (in Mb eV). These data for the most intense resonances are listed in Table~\ref{S+Fits} configuration as well as their main LS character. The MCDF results indicate strong spin-orbit mixing of different LS terms which is indicative of departure from good LS coupling conditions and labelling, hence the repetition of the same leading term for several of the resonances of Table~\ref{S+Fits}. The oscillator strength associated with the $2p^63s^23p^3\rightarrow 2p^53s^23p^3n\text{d}, \ n=4,5,6,7$ transitions is distributed, in the $180-185$ eV photon energy range, over a large number of weak and mixed resonances which do not form the distinct pattern of a Rydberg series. The photon energies corresponding to the $2p^53s^23p^3 {\text{ }}^{3,5}\text{P}$ limits are located in the $182-183$ eV energy gap.
	
	Figure~\ref{S+exptheo} indicates a reasonable agreement between the measured total photoionisation cross sections and those calculated within the MCDF and DARC  theoretical frameworks. This is evidenced by a favourable experiment-theory comparison between both resonance energies and relative intensities and the integrated intensity (in Mb eV) values of 36.7 (Expt), 35.0 (MCDF) and 30.7 (DARC) in the $2p\rightarrow3d/4s$ energy region. The latter two values were deduced from synthetic spectra obtained by summing the theoretical photoionisation cross sections from the $^{4}\text{S}_{3/2}$ ground and $^{2}\text{D}_{3/2}$, $^{2}\text{D}_{5/2}$, $^{2}\text{P}_{1/2}$, $^{2}\text{P}_{3/2}$ metastable states, with weight coefficients of 0.70, 0.11, 0.17, 0.02 and 0.03, and those given in subsection~\ref{darc} for the MCDF and DARC calculations, respectively. 
	
	In the case of the MCDF results of Fig.~\ref{S+exptheo}(b), all the resonances in the 170-190 eV photon range, were given the same Lorentzian width of 0.069 eV. This is the energy equivalent of a decay rate of $1.047\times10^{14} \ \text{s}^{-1}$ which was calculated to be the largest autoionisation rate of the $2p^53s^23p^33d$ configuration via the $2p^53s^23p^33d\rightarrow2p^63s^23p3d\varepsilon \bar e$ spectator Auger transition ($\varepsilon \bar e$ represents the continuum electron). Finally, a convolution with a normalised Gaussian function of FWHM 100 meV to simulate the experimental broadening was applied to generate the spectra of Figs.~\ref{S+exptheo}(b) and (c).The MCDF calculations predict the photon energies of the $2s2p^63s^23p^3 np \  \text{with} \ $ n= 3-7 in the 225-255 eV photon energy range. The $2s2p^63s^23p^4\rightarrow2s^22p^53s^23p^3\varepsilon \bar e$ Coster-Kronig transition has the largest autoionisation rate of the $2s2p^63s^23p^4$ configuration with a value of  $1.6295\times10^{15} \ \text{s}^{-1} \equiv \text{Lorentzian width of} \ 1.073\ \text{eV}$. Photon energies greater than 200 eV were not accessed during the run of \S$^{+}$ experiments and these theoretical predictions could not be checked against experimental data.

\squeezetable	
\begin{table*}
\caption{\label{S+Fits}Comparison of experimental energies (eV) and strengths (Mb eV) with ab initio DARC and MCDF theoretical values for the most intense \S$^{+}$ autoionising resonances belonging to the $2p^53s^23p^33\text{d}$ configuration.}
\begin{ruledtabular}
\begin{tabular}{ccccccc}
\multicolumn{2}{c}{\textrm{Experiment}} &\multicolumn{2}{c}{\textrm{DARC}}& \multicolumn{3}{c}{\textrm{MCDF}} \\
\cline{1-2} \cline{3-4} \cline{5-7}
\vspace{-0.15cm}\\
Photon Energy  & Strength\footnote{The number within parentheses is the uncertainty on the last digit}  & Photon Energy  & Strength  & Photon Energy  & Strength  & Main LS character \\
176.07& 1.566(4) & 178.86 & 0.926 & 175.09 & 1.132 & $^4\text{D}_{5/2} $ \\  
176.22 & 3.689(4) &179.04 &5.653 & 175.32 & 4.455  & $^4\text{D}_{5/2} $ \\
176.36 & 1.718(4) & & &175.39 & 2.359  & $^4\text{D}_{3/2} $ \\
     176.46       &      0.783(3)       &         &                  & 175.46 & 3.021  & $^4\text{F}_{5/2} $ \\
176.60 & 5.996(11)&174.24&7.224&  175.62 & 4.94  & $^4\text{D}_{3/2} $ \\
     176.68&4.126(11)&    179.42   &                4.541                      & 175.65 & 3.514  & $^4\text{P}_{1/2} $ \\
176.85 &  1.109(4)& &    & 175.85 & 0.558  & $^2\text{P}_{1/2} $ \\
176.95 & 1.121(3) & 179.52 & 1.655 & 175.98 & 0.880 & $^2\text{P}_{1/2} $\\
177.19 & 1.783(3) & 179.70 & 3.241 & 175.16 & 2.898 &$^4\text{D}_{5/2} $
\end{tabular}
\end{ruledtabular}
 \end{table*}

	\subsection{\label{Si-like}Silicon  sequence: S$^{2+}$}
\begin{figure}
\includegraphics[width=86mm]{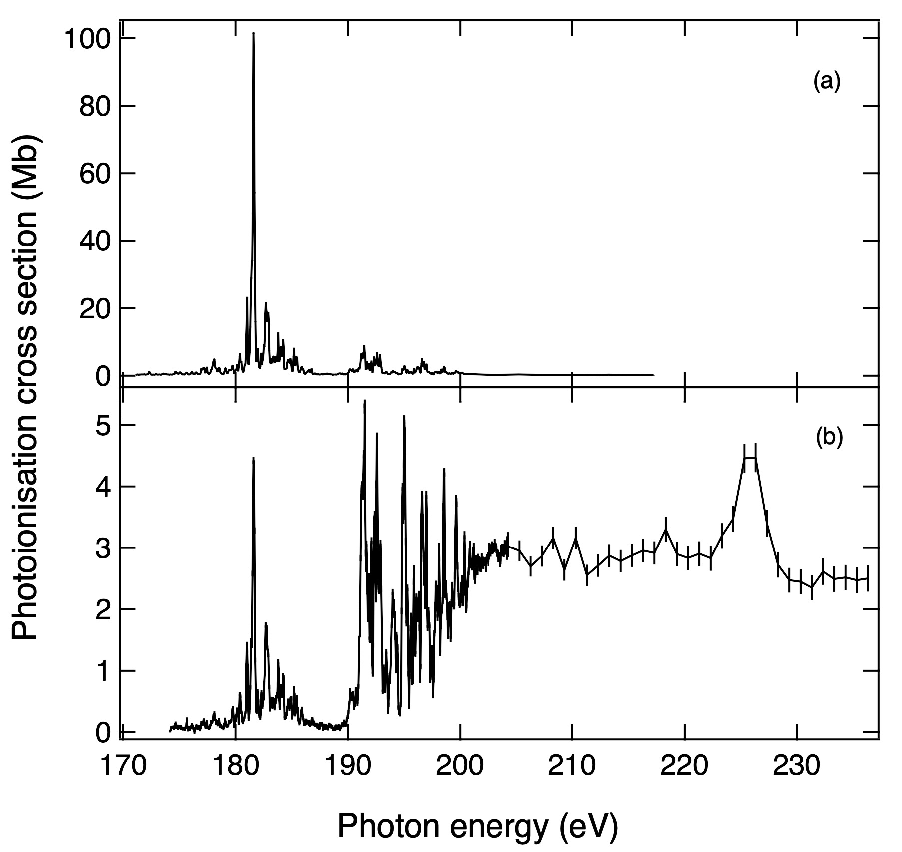}
\caption{\label{S2+exp}Absolute photoionisation cross sections of doubly ionised (\S$^{2+}$) sulphur measured, with a photon energy bandpass of 100 meV, in the (a) single ionisation (SI) and (b) double ionisation (DI) channels, respectively.}
 \end{figure}

 \begin{figure}
\includegraphics[width=86mm]{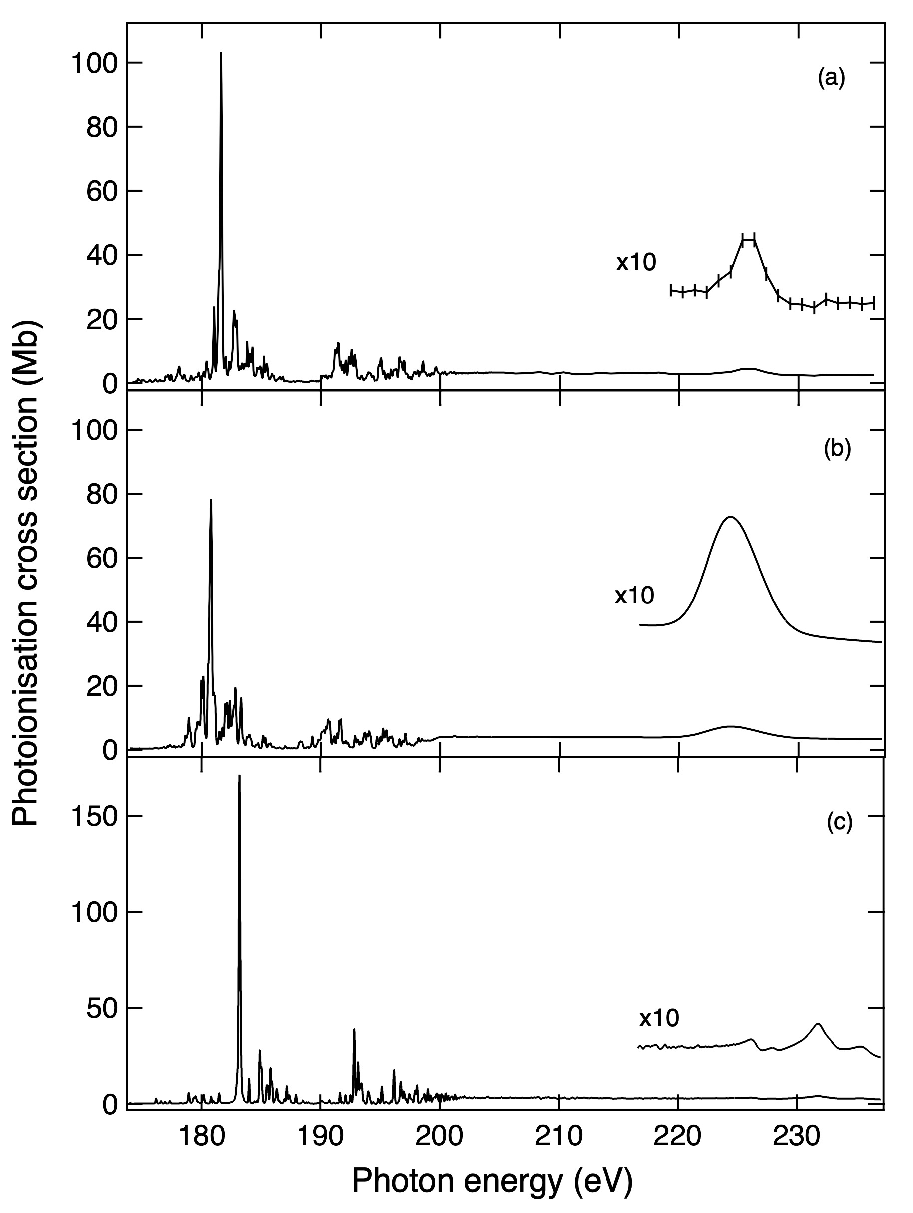}
\caption{\label{S2+exptheo}Total photoionisation cross sections of doubly-Ionised sulphur \S$^{2+}$ in the region of the 2p threshold: (a) Experimental values obtained in this work, (b) multiconfigurational Dirac-Fock (MCDF) theoretical values, (c)  R-Matrix  theoretical values. The MCDF and R-Matrix theoretical cross sections were convolved with normalised Gaussian functions of FWHM 100 meV and 160 meV for photon energies ${<}$215 eV and ${>}$215 eV, respectively, to simulate the experimental broadening of FIG.~\ref{S2+exptheo}(a). The  MCDF and R-Matrix theoretical energies are the ab-initio values. The  MCDF and R-Matrix theoretical intensities were scaled to account for metastable state populations (see text).}
 \end{figure}

	The measured SI and DI cross sections for \S$^{2+}$ are presented in Figures~\ref{S2+exp} (a) and (b), respectively, whilst the total experimental photoionisation cross sections are compared with the MCDF and R-Matrix theoretical results in Figures~\ref{S2+exptheo} (a),(b) and (c). Based on the MCDF results, we can assign the multiple intense resonances observed in the 175-187 eV photon range to the autoionisation decay of many atomic states belonging to the $2p^53s^23p^23\text{d}$ configuration. In this same photon energy range, comparing Figs~\ref{S2+exp} (a) and (b), we measure an intensity ratio of $\text{DI/SI}= 8\%$. It is noticeable that the $2p^53s^23p^23\text{d}$ SI and DI resonance patterns are found at the same peak photon energies with similar relative intensities. This is indicative of the prevalence of correlation type effects, such as shake-off, as the main DI physical mechanism for these resonances.  From figure~\ref{S2+exp}(a), a distinct Rydberg series pattern is discernable with the resonances from the $2p^53s^23p^24\text{d},5\text{d},6\text{d}$  configurations found in the vicinity of the 192 eV, 196.5 eV and 198.5 eV photon energies, respectively. From figure~\ref{S2+exp}(b), the onset of 2p continuous absorption peaks around 204 eV photon energy at the end of a step a few eV wide. This is consistent with the expected structure of multiple thresholds obtained from the 2p$^5$3s$^2$3p$^2$($^3$P) configuration (Table~\ref{Excitation Schemes}). Additional continuum channels may, thus, be open to some of the 2p$^5$3s$^2$3p$^2$nd Rydberg resonances lying in this energy window which would account for the rich resonance pattern oberved in Fig.~\ref{S2+exp}(b). From Fig.~\ref{S2+exp}(b), we note the prominent resonance corresponding to the 2s$^2$3p$^2$($^3$P) $\rightarrow$ 2s2p$^3$ excitations centred near the 225 eV photon energy. The experimental profile appears slightly asymmetric with a measured FWHM of $\sim$ 2.6 eV. Coupling of the 2s($^2$S) spin and angular momenta with those of the three possible terms of the 3p$^3$ ($^4$S, $^2$D, $^2$P) equivalent configuration yields $^3\text{S}_1,\  ^3\text{P}_{0,1,2},\  ^3\text{D}_{1,2,3}$ states and a total of fifteen LS allowed possible resonances from the $2s^22p^63s^2\big(3p^3 \ ^3\text{P}_{0,1,2}\big)$ LSJ ground states. The observed profile is the superposition of all these resonances. Additional strength contributions from the $2s2p^63s^2\big(3p^3 \ ^1\text{D}_{2}\ \text{and}\ ^1\text{S}_0\big)$ metastable states have also to be taken into account (see below) to interpret the breakdown of all the atomic states contributing to the strength and width of the observed wide profile.

	Figure~\ref{S2+exptheo} indicates a good agreement between the measured total photoionisation cross sections and those calculated within the MCDF and R-Matrix theoretical frameworks. This agreement is supported by a favourable experiment-theory comparison between both the resonance energies and relative intensities in the 179-240 eV region. The integrated intensity values (in Mb eV) of 66.5 (Expt), 67.1 (MCDF) and 60.1 (R-Matrix) in the 179-187 eV window (ie. the 2p$\rightarrow$ 3d region) are also in good agreement. Figs~\ref{S2+exptheo}(b) and (c) are synthetic spectra obtained by summing the theoretical photoionisation cross sections from the $^{3}\text{P}$ ground and $^{1}\text{D}_{2}$, $^{1}\text{S}_{0}$ metastable states, with weight coefficients of $\approx$ 0.74, 0.26 and 0.07, respectively. In Fig.~\ref{S2+exptheo}(b), all the resonances below the 2p threshold (ca 205 eV) were given a Lorentzian profile of FWHM 26 meV. This is the energy equivalent of a decay rate of $3.914\times10^{13} \ \text{s}^{-1}$ which was calculated to be the largest autoionisation rate of the $2p^53s^23p^23d$ configuration via the $2p^53s^23p^23d\rightarrow2p^63s^23d\varepsilon \bar e$ spectator Auger transition ($\varepsilon \bar e$ represents the continuum electron).  All the resonances above the 2p threshold (ca 205 eV) were given a Lorentzian profile of FWHM 838 meV. This is the energy equivalent of a decay rate of $1.272\times10^{15} \ \text{s}^{-1}$ which was calculated to be the largest autoionisation rate of the $2s2p^63s^23p^3$ configuration via the $2s2p^63s^23p^3\rightarrow2s^22p^53s^23p^2\varepsilon \bar e$ Coster-Kronig transition ($\varepsilon \bar e$ represents the continuum electron). In those conditions, the synthetic spectrum of Fig.~\ref{S2+exptheo}(b) shows an overall width of $\sim$ 5 eV for the array of 2s$^2$3p$^2$($^3$P) $\rightarrow$ 2s2p$^3$ resonances. This value is somewhat larger than the experimental value of $\sim$ 2.6 eV, the instrumental band pass (BP) of 0.16 eV being neglected here. The MCDF and R-Matrix theoretical cross sections of Fig.~\ref{S2+exptheo}(a) and (b) were convolved with normalised Gaussian functions of FWHM 100 meV and 160 meV for photon energies ${<}$215 eV and ${>}$215 eV, respectively, to simulate the experimental broadening of Fig.~\ref{S2+exptheo}(a).

	 \begin{figure}
\includegraphics[width=86mm]{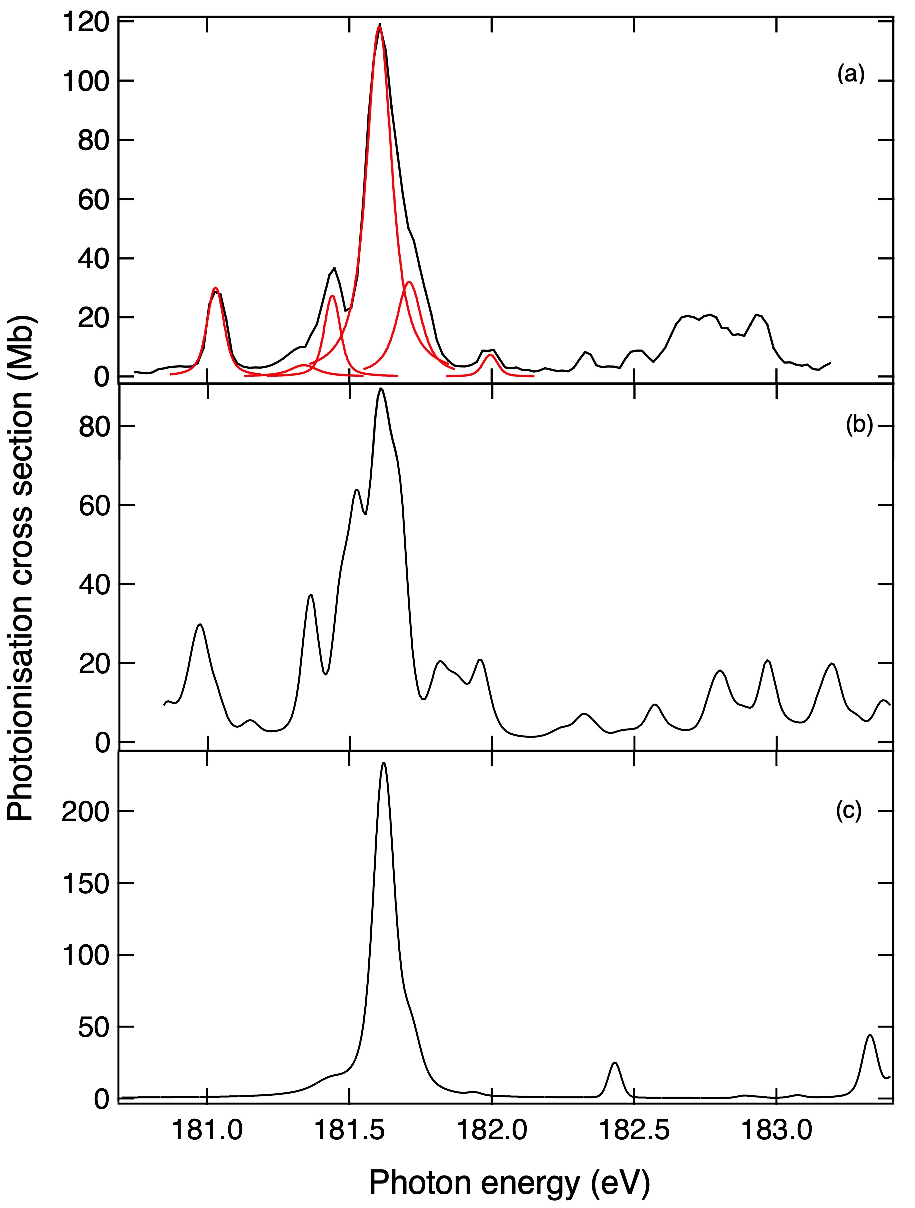}
\caption{\label{S2+exptheohighres}Photoionisation  cross sections of doubly-ionised sulphur \S$^{2+}$ in the photon energy region of the 2p threshold: (a) Single ionisation experimental values obtained in this work at high spectral resolution (BP = 50 meV), (b) multiconfigurational Dirac-Fock (MCDF) total cross sections theoretical values, (c)  R-Matrix  total cross sections  theoretical values. The MCDF and R-Matrix theoretical cross sections were convolved with normalised Gaussian functions of FWHM 50 meV, respectively, and scaled with appropriate metastable population coefficients (see text), to best mimic  the experimental spectrum of FIG.~\ref{S2+exptheohighres}(a).}
 \end{figure}
 
\begin{figure}
\includegraphics[width=86mm]{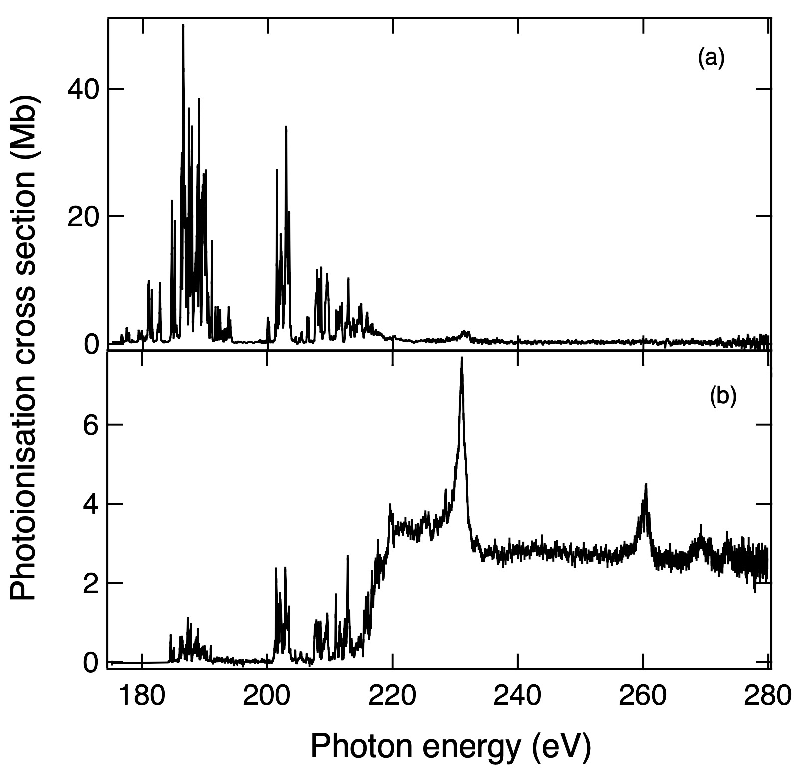}
\caption{\label{S3+exp}Absolute photoionisation cross sections of triply ionised (\S$^{3+}$) sulphur measured, with a photon energy bandpass of 100 meV, in the (a) single ionisation (SI) and (b) double ionisation (DI) channels, respectively.}
 \end{figure}

\begin{figure}
\includegraphics[width=86mm]{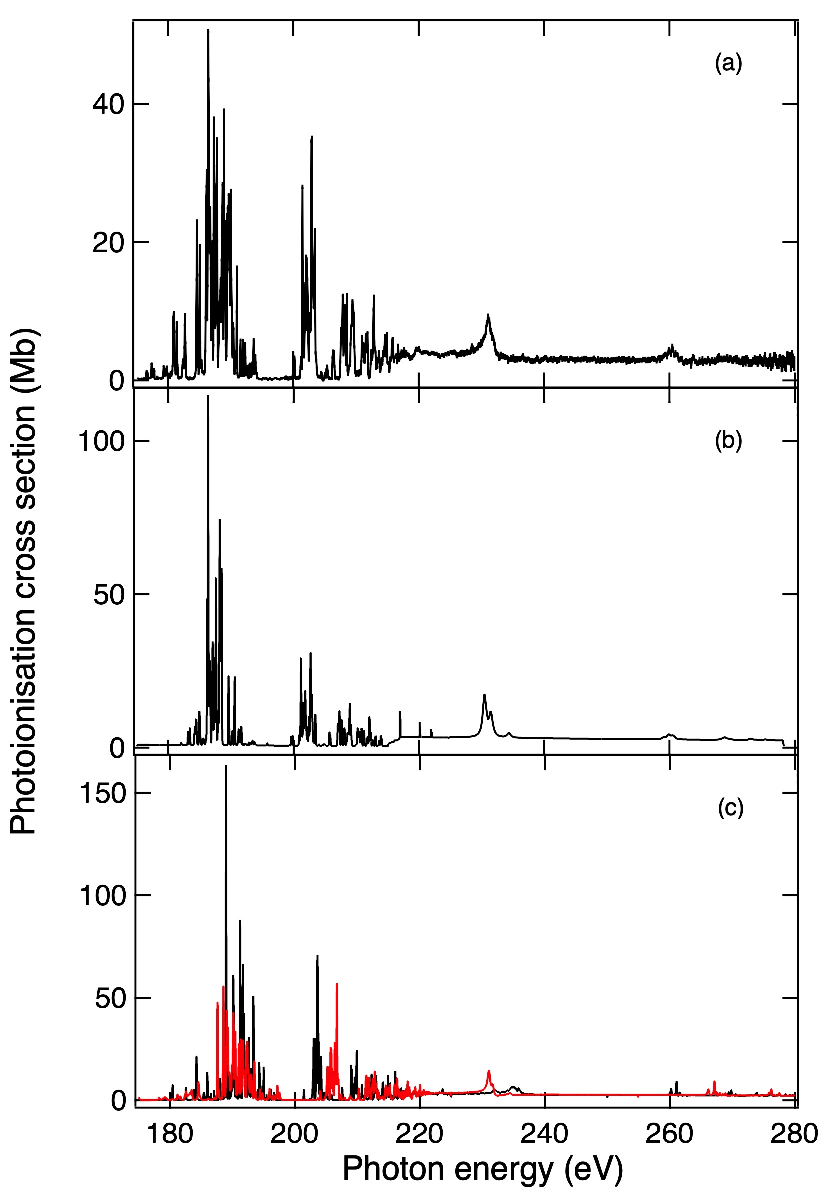}
\caption{\label{S3+exptheo}Total photoionisation cross sections of triply ionised sulphur \S$^{3+}$ in the photon energy region of the 2p threshold: (a) Experimental values obtained in this work, (b) multiconfigurational Dirac-Fock (MCDF) theoretical values, (c)  R-Matrix  LS coupling (black trace) and JK coupling (red trace) theoretical values. The MCDF and R-Matrix theoretical cross sections were convolved with normalised Gaussian functions of FWHM 100 meV, respectively, to simulate the experimental broadening of FIG.~\ref{S3+exptheo}(a)}.
 \end{figure}

\squeezetable
\begin{table*}
\caption{\label{VoigtFitsS2+} Photon energy $E$ (eV), strength $S$ (Mb eV), and natural width $\Gamma\!_L$ (meV) of some \S$^{2+}$ $2p^53s^23p^23\text{d}$ autoionising resonances obtained from the deconvolution of high resolution photon energy scans. The MCDF and R-Matrix ab initio resonance energy values are shifted by +0.84 eV and -1.55 eV, respectively. The theoretical strengths are not corrected for contributions from initial metastable levels populations.}
\begin{ruledtabular}
\begin{tabular}{ccccccccccc}
\multicolumn{3}{c}{\textrm{This work}} & \multicolumn{4}{c}{\textrm{MCDF}} & \multicolumn{4}{c}{\textrm{R-Matrix}}\\
\cline{1-3} \cline{4-7} \cline{8-11}
\vspace{-0.15cm}\\
$E$   & $S$   & $\Gamma\!_L$   & $E$   & $S$   & $\Gamma\!_L$   & $2p\rightarrow 3d$ & $E$   & $S$   & $\Gamma\!_L$   & Initial state \\
\vspace{-0.15cm}\\

181.03 & $2.82\pm0.03$ & $38\pm1$ & 180.98 & 5.74 & 26  & $^{1}\!D_2\rightarrow {^3}\!D_2$ &  &  &  & \\
181.34 & $0.66\pm0.22$ &   & 181.37 & 37.82 & 26 & $^{1}\!S_0\rightarrow {^1}\!P_1$ &  &  &  & \\
181.44 & $2.35\pm0.14$ &   & 181.47 & 5.37 & 26 & $^{3}\!P_2\rightarrow {^5}\!D_3$ & 181.43 & 7.01 & 124 & $^{1}\!D$ \\
 &  &   & 181.53 & 15.59 & 26 & $^{3}\!P_1\rightarrow {^3}\!D_2$ &  &  &  &  \\
181.61 & $18.49\pm0.19$ &  $85\pm1$ & 181.59 & 9.06 & 26 & $^{3}\!P_2\rightarrow {^3}\!D_3$ & 181.61 & 28.48 & 49 & $^{3}\!P$ \\
&  &   & 181.63 & 9.44 & 26 & $^{1}\!D_2\rightarrow {^3}\!F_3$ & 181.65 & 19.62 & 33 & $^{1}\!D$  \\
181.71 & $4.91\pm0.15$ &  $83\pm2$ & 181.68 & 9.21 & 26 & $^{3}\!P_2\rightarrow {^3}\!P_2$ & 181.72 & 5.55 & 53 & $^{3}\!P$ \\
182.00 & $0.57\pm0.02$ & $25\pm3$ & 181.96 & 2.98 & 26 & $^{3}\!P_1\rightarrow {^3}\!P_0$ &  &  &  & \\
\end{tabular}
\end{ruledtabular}
 \end{table*}
 
To enable a more detailed and quantitative comparison between experiment and the theories, a high resolution (BP$=50$ meV) spectrum was recorded over the narrow 181-183 eV photon range. This is shown in figure~\ref{S2+exptheohighres} together with the corresponding MCDF and R-Matrix cross sections. The theoretical curves were energy-shifted by +0.84 eV (MCDF) and -1.55 eV (R-Matrix) to align the most intense peak resonances on the experimental 181.61 eV photon energy value. Applying the relevant metastable population factors, provides integrated strength values of 38.7 Mb eV (MCDF) and 35.0 Mb eV (R-Matrix) to be compared with the experimental value of 38.6 Mb eV.  We were able to perform meaningful deconvolutions of the measured profiles in these high resolution experimental conditions, enabling us to provide estimates for the natural width of the deconvolved resonances. The latter are shown as red traces on Figure~\ref{S2+exptheohighres}.The oscillator strengths and natural widths obtained from this numerical processing of the data are shown in Table~\ref{VoigtFitsS2+}. The detailed agreement between the experimental and MCDF traces of Figs~\ref{S2+exptheohighres} (a) and (b), respectively, is noticeable. In the 181-183 eV window, MCDF predicts at least fifteen resonances, very closely spaced in energy, connecting the GS $2p^63s^23p^2\ ^3\text{P}_{0,1,2}$ components ($0 \mbox{--}1= 37$\ meV and $1\mbox{--}2=103$\ meV \citep{Kramida_2023}) with the numerous 2p$^5$3s$^2$3p$^2$3d $^3$S$_1$, $^3$P$_{0,1,2}$,  $^3$D$_{1,2,3}$ states. The values of some of the measured resonance parameters (strength and natural width) shown in Table~ \ref{VoigtFitsS2+} thus represent $J$-averaged values.

\subsection{\label{Al-like}Aluminium sequence: S$^{3+}$}
	The measured SI and DI cross sections for \S$^{3+}$ are presented in Figures~\ref{S3+exp} (a) and (b), respectively, whilst the total experimental photoionisation cross sections are compared with the MCDF and R-Matrix theoretical results in Figures~\ref{S3+exptheo} (a),(b) and (c). Figure~\ref{S3+exp} (a) shows that the spectrum of SI resonances is organised into clearly separated photon energy groups, each containing many narrow resonances, of marked decreasing intensity as the photon energy increases. The first three groups observed in the 182-190 eV, 200-203 eV and $\sim$207-208 eV are unambiguously attributable to the non radiative decay of the $2p^53s^23p3d/4s,\,2p^53s^23p4d/5s \, \text{and} \, 2p^53s^23p5d/6s$ configurations, respectively. Reasonably accurate measurement of resonance energies and strengths can be obtained from profile fitting of the experimental spectra in the $2p\rightarrow3d/4s$ and $2p\rightarrow4d/5s$ regions. The data is presented in Table~\ref{S3+Table} together with detailed comparisons with the MCDF and R-Matrix (JK coupling) results. These comparisons are based on visually best-matching the observed and calculated resonance energies and intensities and are not the results of computer-based protocols. The 3d/4s and 4d/5s SI resonance energy groups are retrieved in the DI spectrum of Figure~\ref{S3+exp} (b), with the 4d/5s group being overall more intense than the 3d/4s group. The DI mechanism for these resonances below the 2p threshold is similar to the case of S$^{2+}$ discussed in sub-section \ref{Si-like}. From Figure~\ref{S3+exp} (b), the $2p^53s^23p\ ^{1,3}\text{S}\ ^{1,3}\text{P}\ ^{1,3}\text{D}$ threshold at 219.0 eV and width of 1.74 eV rises more sharply than its  S$^{2+}$ counterpart (see Fig.~\ref{S2+exp} (b)). The series of  $2s2p^63s^23pnp,\ n\geq 3$ resonances feature prominently in the DI spectrum of Figure~\ref{S3+exp} (b) ($2s2p^63s^23p^2\sim231\ \text{eV},\ 2s2p^63s^23p24p \sim261\ \text{eV},\ 2s2p^63s^23p5p\sim270\ \text{eV}$). The $2s2p^63s^23p^2\sim231\ \text{eV}$ resonance has a visible counterpart ($\sim$ 2 Mb peak) in the SI cross section spectrum. We propose the following mechanisms to interpret these data. For the DI process, the $2s$ vacancy decays via any of the possible $L_{1}\!-\!L_{2,3}M_y$ Coster-Kronig processes. This leaves the S$^{4+}$ ion with a single 2p vacancy that will then decay via $L_{2,3}\!-\!M_xM_y$ Auger processes. The final S$^{5+}$ ion is now in a stable configuration. For the SI process, the $2s$ vacancy decays via any of the possible $L_{1}\!-\!M_xM_y$ resonant Auger processes, leaving the S$^{4+}$ ion in either the stable $2s^22p^63s^2$ configuration or valence-excited configurations that are themselves either metastable ($2s^22p^63p^2$) or allowed ($2s^22p^63s3p$) against radiative decay. We note that the $2s2p^63s^23p^2$ DI resonance feature at $\sim$ 231 eV exhibits the characteristically asymmetric Fano profile indicating non-negligible interactions with the underlying continua.
	
The experimental total photoionisation cross sections are presented and compared with the ab initio (i.e. no photon energy shifts nor population intensity factors were applied) MCDF and R-Matrix theoretical ones on Fig.~\ref{S3+exptheo} (a), (b) and (c), respectively. The theoretical cross sections were convolved with Gaussian profiles of width 100 meV and 160 meV, for the photon energies less or greater than 220 eV, respectively. The agreement between the theories and the experiment appears very satisfactory in terms of reproducing correctly the main experimental  spectral features as well as their relative intensities. The integrated cross sections for the 3d/4s energy group are 102.7, 98.0 and 100.5 Mb eV for the experiment, MCDF and R-Matrix (JK coupling) calculations, respectively. The MCDF cross sections presented in Fig.~\ref{S3+exptheo}(b) are "dressed" with normalised Lorentzian profiles of FWHM 7 meV and 582 meV for the resonances below and above the 2p threshold, respectively. These correspond to calculated autoionisation rates of $1.114\times10^{13}\ \text{s}^{-1}$ and $8.845\times10^{14}\ \text{s}^{-1}$ for the $2p^53s^23p3d\ \rightarrow \ 2p^63s3d$ and $2s2p^63s^23p^2\ \rightarrow \ 2s^22p^53s^23p$ spectator Auger and Coster-Kronig transitions, respectively. 

MCDF interprets the single resonance observed at 231.2 eV as the blend of the total of nine possible $LSJ$ allowed transitions $2s^22p^63s^23p\ ^{2}\text{P}_{1/2,3/2}\rightarrow 2s2p^63s^23p^2(^3\text{P})\ ^{2}\text{P}_{1/2,3/2}$, $\ 2s3p^2(^1\text{D}) ^{2}\text{D}_{3/2,5/2}, \ 2s3p^2(^1\text{S}) ^{2}\text{S}_{1/2}$, ie the $2s^22p^63s^23p\,  ^{2}\text{P} \rightarrow 2s2p^63s^23p^2\, ^2\text{P},^2\text{D},^2\text{S}$ multiplet. The energy and strength details for each of the nine resonances are given in Table~\ref{S3+Table}. As the experimentally observed cross section profile is that of the (unresolved) multiplet, it can be parametrised following the approach described in \citep{FanoCooper_1968} and, thus compared, with the multiplet averaged values predicted by theory. The Fano Cooper parametrisation, using the notation given in \citep{FanoCooper_1968} gives the following values (the BP of 0.1 eV is neglected throughout): $\sigma_c=3.04$ Mb, $\rho^2=7.24\times10^{-2}$, $\Gamma_R=1.37$ eV, $E_R=231.19$ eV and $q=-4.24$. The $\sigma_c(\rho^2q^2)/(1+\epsilon^2)$ profile, with $\epsilon=(E-E_R)/\frac{1}{2}\Gamma_R$ the reduced energy, represents the intrinsic, Lorentz shaped, profile of an autoionising line, unaffected by interference (refer to \citep{FanoCooper_1968} for details) and would, thus, be directly comparable with the MCDF predictions. The integration over energy of this profile gives an oscillator strength of $\sim 0.078 $ which compares favourably with the R-Matrix value of $\sim 0.09 $ obtained from numerical integration of figure~\ref{S3+exptheo}(c) and therefore comprises the effect of spectral repulsion in addition to the Lorentzian contribution. The agreement with the MCDF multiplet averaged value of 0.19 appears somewhat less satisfactory being twice as large as the experimental value.

\squeezetable
\begin{table*}
\caption{\label{S3+Table} Photon energy $E$ (eV), strength $S$ (Mb eV), and natural width $\Gamma\!_L$ (meV) of some \S$^{3+}$ $2p^53s^23p3\text{d},\ 2p^53s^23p4\text{d}\ \text{and}\ 2s2p^63s^23p^2$ autoionising resonances. The MCDF and R-Matrix ab initio resonance energy values are unshifted. The theoretical strengths are not corrected for contributions from initial metastable levels populations.}
\begin{ruledtabular}
\begin{tabular}{ccccccccc}
\multicolumn{2}{c}{This work} & \multicolumn{3}{c}{MCDF} & \multicolumn{4}{c}{R-Matrix (JK coupling)}\\
\cline{1-2} \cline{3-5} \cline{6-9}
\vspace{-0.10cm}\\
$E$ \footnotemark[1]   & $S$ \footnotemark[1]   & $E$   &  $S$  & ${^2}P_{1/2,3/2} \rightarrow{^{2S+1}}\negmedspace L_J$ & $E$    & $S$   & $\Gamma\!_L$  & $^2P_{1/2,3/2}$ Initial State \\
\vspace{-0.15cm}\\
\multicolumn{9}{l}{$2p\rightarrow3d$ \textit{excitations}}\\
180.916(2) & 1.02(2) &  &   &    &     &    &   &   \\
184.631(1)  & 2.34(3) & 184.94 & 1.63  & ${3/2} \rightarrow {^2}\! P_{3/2} $   &     &    &   &   \\
185.112(1) & 1.85(4) &  &   &    &     &    &   &   \\
186.048(1) & 1.15(3) &  &   &    &     &    &   &   \\
186.143(1) & 2.44(3) &  &   &    &     &    &   &   \\
186.233(1) & 2.77(3) & 186.20 &  7.38 &  ${3/2} \rightarrow {^2}\! D_{5/2} $   &    &    &   &   \\
186.425(1) & 5.21(2) & 186.34 & 13.40  &  ${3/2} \rightarrow {^4}\! D_{5/2} $  &  187.63   &  8.99  & 20  & $3/2$   \\
186.525(1) & 4.27(3) & 186.37 &  13.08 &  ${1/2} \rightarrow {^2}\! D_{3/2} $  &     &    &   &   \\
186.645(1) & 2.22(2) & 186.60 & 7.83  & ${1/2} \rightarrow {^4}\! D_{3/2} $    &     &    &   &   \\
 186.772(0) & 2.97(2)&   186.70  &   4.85    &   ${1/2} \rightarrow {^4}\! D_{1/2} $     &           &          &         &          \\
 187.025(2)&   1.05(2)  &  186.97     &   2.56     &     ${3/2} \rightarrow {^4}\! D_{3/2} $      &     188.46     &   1.82      &  11    &   3/2     \\
 187.128(1) &  2.15(2)   &   187.07    &  3.14      &     ${3/2} \rightarrow {^4}\! P_{1/2} $      &          &         &      &        \\
 187.344(0) &  3.53(2)   &  187.36     &   9.52     &    ${1/2} \rightarrow {^4}\! D_{3/2} $  &  188.58   &     17.6     &     11    &    1/2   \\
 187.459(0) &  2.94(2)   &       &        &           &          &         &      &        \\
187.612(1)  &  1.00(2)   &       &        &           &          &         &      &        \\
187.755(1)  &  3.07(2)   &   187.62    &    7.95    &    ${3/2} \rightarrow {^2}\! D_{5/2} $       &    189.11      &    8.78     &  47    &    3/2    \\
 187.863(1)  &  3.53(2)   &   188.20    &    12.4    &    ${3/2} \rightarrow {^2}\! D_{3/2} $       &    189.26      &    7.75     &  59    &    3/2        \\
 188.294(1)  &  1.25(2)   &   188.32    &   2.97    &    ${3/2} \rightarrow {^2}\! D_{5/2} $       &    189.53      &    3.68     &  39    &    1/2          \\
 188.418(1)  &  1.61(2)   &   188.39    &   3.06    &    ${3/2} \rightarrow {^2}\! P_{1/2} $       &    189.81      &    3.31     &  7    &    1/2            \\
188.574(1)  &  1.77(2)   &   188.41    &   2.66    &    ${3/2} \rightarrow {^2}\! D_{3/2} $       &    190.12      &    2.90     &  4    &    3/2             \\
188.684(1)  &  1.97(3)   &   188.50    &   8.75    &    ${1/2} \rightarrow {^2}\! P_{1/2} $           &          &         &      &        \\
188.778(1)  &  3.06(3)   &   188.52    &   10.8    &    ${1/2} \rightarrow {^2}\! D_{3/2} $           &     190.19     &    5.84     &   11   &      3/2    \\
188.966(0)  &  4.40(2)   &       &      &             &     190.48     &   12.3     &   20   &      1/2    \\
189.093(1) & 1.93(5) &  &   &    &     &    &   &  \\
189.164(2) & 1.55(5) &  &   &    &     &    &   &   \\
189.472(1) & 2.66(2) &  &   &    &  190.91   &   1.73 & 4  & 3/2  \\
189.589(1) & 2.87(2) &  &   &    &  191.04   &   2.74 & 2  & 3/2  \\
189.726(0)  &  3.30(2)   &   189.61    &   3.78    &    ${3/2} \rightarrow {^2}\! S_{1/2} $           &     191.33     &    5.28     &   17   &      3/2    \\
189.855(1) & 1.93(2) &  &   &    &     &    &   &  \\
190.018(1) & 2.94(2) &  &   &    &  191.59   &  4.33  &  50 & 3/2 \\
190.125(1) & 1.71(2) &  &   &    &  191.69   &  3.87  &  43 & 3/2 \\
190.441(1)  &  0.98(1)   &   190.59    &   3.80    &    ${3/2} \rightarrow {^2}\! D_{5/2} $           &     191.91     &    2.36    &   30   &      3/2    \\
191.020(1) & 1.50(2) &  &   &    &  192.28   &  4.41  &  17 & 3/2 \\
\vspace{-0.15cm}\\
\multicolumn{9}{l}{$2p\rightarrow4d$ \textit{excitations}}\\
201.455(1) & 2.90(4) & 201.56 &  1.94 &  ${3/2} \rightarrow {^2}\! D_{5/2} $   &    &    &   &   \\
201.560(2) & 1.42(3) & 201.71 &  1.55 &  ${3/2} \rightarrow {^2}\! D_{3/2} $   &    &    &   &   \\
201.927(2) & 1.58(3) & 201.82 &  2.83 &  ${1/2} \rightarrow {^2}\! D_{3/2} $   &    &    &   &   \\
202.075(2)  &  1.846(6)   &   201.88    &   1.65    &    ${3/2} \rightarrow {^2}\! P_{3/2} $           &     205.82     &    5.08     &   4   &      1/2    \\
202.168(3) & 1.54(5) &  &   &    &     &    &   &  \\
202.270(2) & 1.48(4) & 202.43 &  2.95 &  ${1/2} \rightarrow {^4}\! F_{3/2} $   &    &    &   &   \\
202.891(2) & 3.21(4) & 202.72 &  2.30 &  ${3/2} \rightarrow {^2}\! D_{5/2} $   &    &    &   &   \\
202.982(1) & 3.27(3) &  &   &    &  206.38   &  8.66  & 6  & 1/2  \\
203.117(1) & 1.89(2) &  &   &    &  206.67   &  4.49  & 4  & 1/2  \\
203.228(1) & 1.33(3) &  &   &    &     &    &   &  \\
203.46(1)  &  2.30(1)   &   203.47    &   1.54    &    ${3/2} \rightarrow {^2}\! F_{5/2} $           &     206.73     &    2.24    &   5   &      3/2    \\
203.35(1) & 1.42(3) &  &   &    &     &    &   &  \\
\vspace{-0.15cm}\\
\multicolumn{9}{l}{$2s\rightarrow3p$ \textit{excitations}}\\
231.2(1) &     &   230.53    &    2.22    &    ${3/2} \rightarrow {^2}\! P_{1/2} $       &         &       &      &        \\
  &     &   230.62    &    10.96    &    ${3/2} \rightarrow {^2}\! P_{3/2} $       &         &       &      &        \\
   &     &   230.64    &    9.23    &    ${1/2} \rightarrow {^2}\! P_{1/2} $       &         &       &      &        \\
    &     &   230.74    &    5.24    &    ${1/2} \rightarrow {^2}\! P_{3/2} $       &         &       &      &        \\
     &     &   231.61    &    6.83    &    ${3/2} \rightarrow {^2}\! D_{5/2} $       &    231.1     &       &      &   3/2     \\
     &     &   231.61    &    1.13    &    ${3/2} \rightarrow {^2}\! D_{3/2} $       &         &       &      &   \\
      &     &   231.73    &    6.74    &    ${1/2} \rightarrow {^2}\! D_{3/2} $       &   231.2      &       &      &    1/2    \\
      &     &   234.47   &    1.38   &    ${1/2} \rightarrow {^2}\! S_{1/2} $       &         &       &      &      \\
      &     &   234.59    &    1.55    &    ${3/2} \rightarrow {^2}\! S_{1/2} $       &       &       &      &    
\footnotetext[1]{The number within parentheses is the uncertainty on the last digit}
\end{tabular}
\end{ruledtabular}
\end{table*}


 \begin{figure}
\includegraphics[width=86mm]{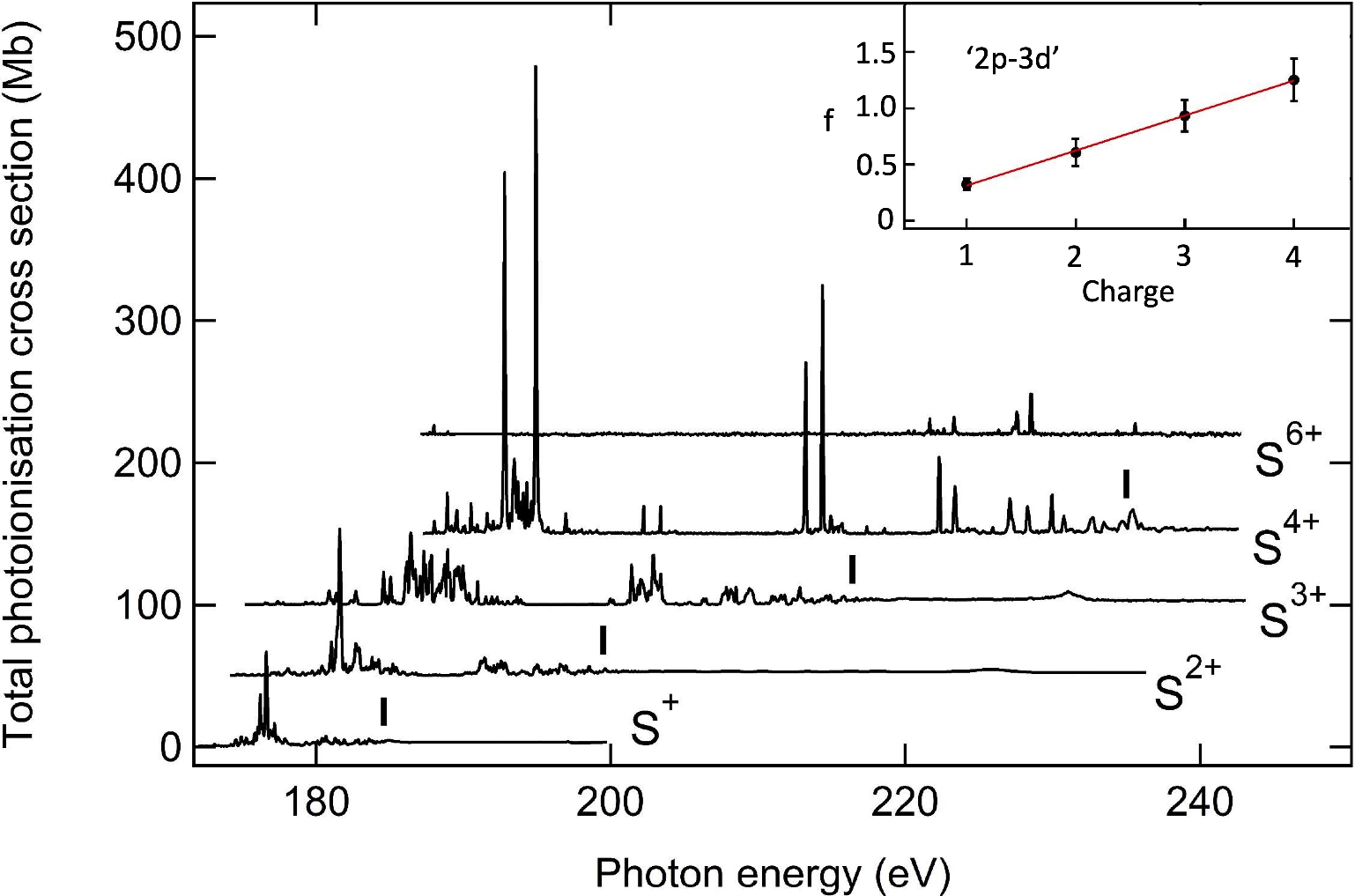}
\caption{\label{isonuclear}Total experimental photoionisation cross sections near the 2p threshold (indicated by a short vertical bar) for the isonuclear series \S$^+$, \S$^{2+}$, \S$^{3+}$, \S$^{4+}$, \S$^{6+}$. The insert shows the total $2p\rightarrow 3d$ oscillator strength obtained from the integration of these plots (except \S$^{6+}$), over the $2p\rightarrow 3d$ photon energy range, as a function of the ionic charge.}
 \end{figure}
\subsection{\label{IsonuclearComp}Isonuclear comparison of \S$^+$, \S$^{2+}$, \S$^{3+}$, \S$^{4+}$, \S$^{6+}$}

In Figure \ref{isonuclear}, we present the behaviour of the total photoionisation cross sections for the ions of the  \S$^+$, \S$^{2+}$, \S$^{3+}$, \S$^{4+}$ isonuclear series, in the region of their respective 2p thresholds, together with that of \S$^{6+}$ in the same energy region. The $2p^5\ ^2P_{1/2,3/2}$ ionisation thresholds of the Ne-like \S$^{6+}$ ion have energies of 281.0 and 282.2 eV \citep{Kramida_2023}, respectively. Thus, the data shown on Figure~\ref{isonuclear} for the \S$^{6+}$ ion is not direct photoionisation from the $2p^6\ ^1S_{0}$ GS, but rather from $2p^53s$ excited states, similar to the previous case of \Si$^{4+}$ \citep{Bizau_2009}. The \S$^{4+}$ data shown on Figure~\ref{isonuclear} were obtained with the help of the same experimental set up as used in the present work and have been published separately in \citep{Mosnier_2022}. The $2p\rightarrow 3d$ oscillator strength ($\equiv$ f$_{2p\rightarrow 3d}$) obtained from the integration of the \S$^+$, \S$^{2+}$, \S$^{3+}$, and \S$^{4+}$ total cross section plots, over the $2p\rightarrow 3d$ photon energy range, is shown in the insert of Figure~\ref{isonuclear} as a function of the ionic charge (1 for \S$^+$, 2 for \S$^{2+}$, etc...). The f$_{2p\rightarrow 3d}$ is seen to increase linearly with increasing ionic charge with a slope of $\approx$ 0.25 per unit charge. Increased configuration interaction effects engaging more electron subshells \citep{Costello_1992} together with changing electron screening effects as the ionicity increases are likely causes for the observed behaviour. Since the f$_{2p\rightarrow 3d}=0.696$ \citep{Wiese_2009} for the last member of the isonuclear sequence, i.e. the H-like \S$^{15+}$ ion, then the complete graph for the sequence must possess a maximum. We note that for the filled 2p subshell of the neon-like \S$^{6+}$ ion, the $9.40\times10^{11}\, \text{s}^{-1}$ radiative rate of the non-autoionising $2p^53d \, ^1\text{P}_1 \rightarrow 2p^6 \, ^1\text{S}_0$ transition at 206.09 eV \citep{Kramida_2023} converts to a value of f$_{2p\rightarrow 3d}=1.53$ which one would expect to be close to the maximum value.

\section{\label{conclusions}Conclusions}
The absolute photoionisation cross sections of the \S$^+$, \S$^{2+}$, \S$^{3+}$ ions have been measured in the region of their 2p$^{-1}$/2s$^{-1}$ thresholds using the merged-beam technique on the MAIA apparatus at the SOLEIL synchrotron radiation facility. Rich and complex resonant structures arising primarily from the decay of 2p excited configurations were observed for all three spectra up to and including the photon energies of the 2p thresholds. With three open subshells the calculation of such resonance series presents a considerable challenge to atomic theories. 

Extensive multiconfiguration Dirac-Fock, Breit-Pauli R-Matrix and Dirac R-Matrix theoretical cross section calculations were carried out.
Our results show that the MCDF and R-Matrix (RMAT) approaches both provide useful insight into
the interpretation of the experimental results. Both theoretical approaches show
good agreement in terms of the integrated cross section (oscillator strength) for the L-excitation region.
MCDF shows quite good detailed agreement with the experimental 2p
resonances while RMAT provides better insight into the
strongly autoionising 2s excitations. In order to satisfactorily model the experimental
data it was necessary to include many coupled states and channels in the
calculations, for both the ground and low-lying metastable states.

To compare the theoretical data with the experimental results, systematic energy
shifts were required. Systematic energy shifts between the predictions of theoretical
resonance positions and experimental results are not unexpected, see e.g., \citep{Mosnier_2023}.
 In either the MCDF or the R-matrix
calculations, the calculated resonant photon energies equal the difference between
the computed quasi-bound, or resonance, state and the computed initial ground
state. These absolute energies in either the MCDF or RMAT calculations are subject
to the variational principle, in that they are an overestimate for any incomplete basis
description. A particular calculation can either correlate the initial state more than
the resonant quasi-bound state, as is the case with the RMAT calculations, leading
to an overestimate of the true photon energy, or correlate the initial state less
compared to the resonant state, giving an underestimate of the true photon energy. 
The R-matrix method is particularly suited for allowing large configuration
expansions for the single initial ground state, hence the difference between RMAT
and MCDF.  Why one converges either the ground or resonance state more or less
is a matter of the "in-completeness" of the basis set for both states and is not easy to
quantify. The existence of such systematic shifts further emphasises the importance
of experimental benchmarking for even the most sophisticated theoretical models.
 
 \begin{acknowledgments}
The authors thank the SOLEIL beamline staff John Bozek and Aleksandar Milosavljevic for their help during the experiments.
TWG was supported in part by NASA (NNX11AF32G).
BMCL acknowledges Queen's University Belfast for a visiting research 
fellowship (VRF). The authors gratefully acknowledge the 
Gauss Centre for Supercomputing e.V. (https://www.gauss-centre.eu/) 
for funding this project by providing 
computing time on the GCS Supercomputer HAZEL HEN at 
H\"{o}chstleistungsrechenzentrum Stuttgart (https://www.hlrs.de/).
\end{acknowledgments}
                                             

\bibliography{Sulphur_ions_JPhysB}
\end{document}